\documentclass[11pt,a4paper]{article}
\pdfoutput=1
\usepackage{jcappub}
\usepackage{rotating}
\usepackage{array}
\usepackage{amsmath}
\usepackage[normalem]{ulem}
\usepackage{slashed}
\usepackage{booktabs}
\usepackage[pdftex,table]{xcolor}
\usepackage{units}
\usepackage{xspace}
\usepackage{makecell}

\newcolumntype{C}[1]{>{\centering\let\newline\\\arraybackslash\hspace{0pt}}m{#1}}

\newcommand{\ba}[1]{\ensuremath{\left( #1 \right)}}
\newcommand{\bb}[1]{\ensuremath{\left[ #1 \right]}}
\newcommand{\bc}[1]{\ensuremath{\left\{ #1 \right\}}}

\newcommand{\nocontentsline}[3]{}
\newcommand{\tocless}[2]{\bgroup\let\addcontentsline=\nocontentsline#1{#2}\egroup}

\newcommand{\plane}[2]{$#1$\,--\,$#2$}
\newcommand*\diff{\mathop{}\!\mathrm{d}}

\newcommand{\fphi}{f_\phi}

\newcommand{\perc}{\ensuremath{\text{p}}}
\newcommand{\peak}{\ensuremath{\text{peak}}}

\newcommand{\GW}{\ensuremath{\text{GW}}}

\newcommand{\ds}{{\sf DarkSUSY}}

\usepackage{multirow}
\usepackage{caption}
\usepackage{subcaption}

\arxivnumber{P3H-25-012, TTP25-004}

\title{Sub-GeV dark matter and nano-Hertz gravitational waves from a classically conformal dark sector}

\author[a]{Sowmiya Balan,}
\author[b]{Torsten Bringmann,}
\author[a,c]{Felix Kahlhoefer,}
\author[a]{Jonas Matuszak,}
\author[d,e]{and Carlo Tasillo}

\affiliation[a]{Institute for Theoretical Particle Physics (TTP), Karlsruhe Institute of
  Technology (KIT), 76128 Karlsruhe, Germany}
\affiliation[b]{Department of Physics, University of Oslo, Box 1048, N-0316 Oslo, Norway}
\affiliation[c]{Institute for Astroparticle Physics (IAP), Karlsruhe Institute of
  Technology (KIT), 76128 Karlsruhe, Germany}
\affiliation[d]{Deutsches Elektronen-Synchrotron DESY, Notkestr.~85, 22607 Hamburg, Germany}
\affiliation[e]{Department of Physics and Astronomy, Uppsala University, Box 516, SE-751 20 Uppsala, Sweden}

\emailAdd{sowmiya.balan@kit.edu, torsten.bringmann@fys.uio.no, kahlhoefer@kit.edu, jonas.matuszak@kit.edu, carlo.tasillo@physics.uu.se}

\abstract{
Strong first-order phase transitions in a dark sector offer a compelling explanation for the stochastic gravitational wave 
background in the nano-Hertz range recently detected by pulsar timing arrays (PTAs). We explore the possibility that such a 
phase transition at the same time gives mass to a stable fermion that accounts for the observed dark matter abundance 
and leads to testable effects in laboratory experiments. Concretely, we consider a classically conformal dark sector with a 
hidden $U(1)^\prime$ gauge symmetry that couples to the Standard Model via kinetic mixing. 
Since the PTA signal requires a phase transition in the MeV temperature range, spontaneous symmetry breaking  gives rise 
to a sub-GeV dark matter candidate that couples to the Standard Model via a dark photon mediator and obtains its relic 
abundance via annihilations into electrons and
dark Higgs bosons. Such a scenario is tightly constrained by laboratory searches for dark 
photons and cosmological constraints on the decays of dark Higgs bosons after the phase transition. We show that 
viable parameter regions can be found both for the case that the dark Higgs bosons remain in equilibrium with the 
Standard Model and that they decouple and only decay much later. In the latter case, the parameter regions preferred by 
the PTA signal and the dark matter relic abundance can be fully explored by future beam-dump experiments searching for 
missing energy.}

\keywords{primordial gravitational waves (theory), cosmology of theories beyond the SM,
  particle physics -- cosmology connection}

\begin{document}
\maketitle
\flushbottom

\section{Introduction}
\label{sec:introduction}

So far, all evidence for the existence of dark matter (DM) is based on its gravitational effects, while any other form of 
interactions -- if present -- has evaded detection. This raises the question whether we can study the fundamental properties 
of DM particles using gravitational interactions alone. A particularly promising avenue are stochastic gravitational wave 
(GW) signals that arise from a strong first-order phase transition (FOPT) involving a scalar field coupled to the DM 
particle~\cite{Alanne:2014bra,Schwaller:2015tja,Baldes:2017rcu,Hall:2019rld}. Analogous to the Higgs mechanism in the 
Standard Model (SM), the vacuum expectation value (vev) of such a scalar field would then generate the mass of the DM 
particle, thus providing a direct link between GW signals and DM 
properties~\cite{Marfatia:2020bcs,Wang:2022akn,Kanemura:2023jiw,Bringmann:2023iuz}.

The recent observation of a stochastic GW background in the nano-Hertz frequency range by
various pulsar timing array (PTA) collaborations, namely the North American Nanohertz
Observatory for Gravitational Waves (NANOGrav)~\cite{NANOGrav:2023hfp}, the European PTA
(EPTA) as well as the Indian PTA (InPTA)~\cite{EPTA:2023fyk}, the Chinese PTA
(CPTA)~\cite{Xu:2023wog}, the Parkes PTA (PPTA)~\cite{Reardon:2023gzh}, the MeerKAT
PTA~\cite{Miles:2024seg} and the International PTA (IPTA)~\cite{Antoniadis:2022pcn}, has
created a significant amount of interest in possible astrophysical origins, as well as
cosmological sources involving physics beyond the standard model (SM) of particle physics.
Among the former are in particular supermassive black hole (SMBH) binaries, while the
latter possibilities encompass inflation~\cite{Vagnozzi:2023lwo, Dandoy:2023jot}, cosmic
strings~\cite{Ellis:2023tsl, Buchmuller:2023aus} and
FOPTs~\cite{Ashoorioon:2022raz,Bringmann:2023opz,Madge:2023dxc,
  Gouttenoire:2023bqy,Banik:2024zwj, Goncalves:2025uwh, Costa:2025csj, Li:2025nja,
  Li:2025rrs, Athron:2023mer}. Many works have studied these different sources and
discussed the possible implications~\cite{NANOGrav:2023hvm,Ellis:2023oxs,
  Figueroa:2023zhu,Guo:2023hyp, Depta:2023qst,Gouttenoire:2023ftk, Franciolini:2023wjm}.
While the observed PTA signal could plausibly originate from SMBH binaries, recent
findings by the NANOGrav collaboration~\cite{NANOGrav:2023hfp} and $N$-body
simulations~\cite{Chen:2025wel} indicate a significant tension between the best fit to the
PTA data and the expected astrophysical parameter ranges for this scenario. This
discrepancy motivates us to hypothesise that at least part of the observed signal may be
due to a FOPT.

The possibility of accounting for the PTA signal through a delayed electroweak phase transition was shown to be unfeasible 
in ref.~\cite{Athron:2023mer}. If the PTA signal is instead interpreted in terms of a FOPT in the dark sector, one can directly 
translate its frequency to the temperature of the FOPT~\cite{Kamionkowski:1993fg}, which for nHz frequencies lies in the 
MeV range.  Since the temperature 
is closely related to the vev of the scalar field, and hence the mass of the dark sector particles, one therefore expects new 
particles below the GeV scale. This mass region has received a rapidly growing amount of interest in recent years for an 
entirely different reason, namely in the context of DM physics. The reason is that it is possible for sub-GeV DM particles to 
maintain the successes of thermal freeze-out, 
while evading the many strong experimental constraints on heavier thermal 
relics, in particular from direct detection experiments~\cite{Arcadi:2017kky,Arcadi:2024ukq}. 

As a result, recent years have brought rapid progress in the experimental programme to search for sub-GeV dark sectors at 
high-intensity beam-dump experiments and low-threshold scattering 
detectors~\cite{Alexander:2016aln,Knapen:2017xzo,Battaglieri:2017aum}. Moreover, particles in this mass range can have 
a wide range of astrophysical and cosmological implications, including signatures of DM 
annihilations~\cite{Slatyer:2009yq,Cirelli:2023tnx}, modifications of the expansion history of the 
universe~\cite{Depta:2019lbe,Sabti:2019mhn}, and effects on the formation and evolution of DM 
halos~\cite{Robertson:2016xjh,Tulin:2017ara}. A recent study~\cite{Balan:2024cmq} used the {\sf GAMBIT} global fitting 
framework~\cite{GAMBIT:2017yxo} to perform a global fit of all of these constraints for scalar and fermionic sub-GeV DM 
candidates interacting with SM particles via the exchange of a new $U(1)'$ gauge boson that kinetically mixes with photons 
(called dark photon). The study identified large viable regions of parameter space that can be targeted by future 
experiments. However, no mechanism for generating the DM and dark photon masses was specified, even though the 
presence of an additional scalar field may significantly change the phenomenology~\cite{Duerr:2016tmh}.

In the present work we connect the idea of sub-GeV DM to the PTA signal by considering a
dark Higgs boson which can obtain a vev that spontaneously breaks the gauge symmetry and
generates particle masses, see also ref.~\cite{Li:2025nja}. We assume furthermore that the
dark sector has an approximate conformal symmetry, i.e.\ that there are no dimensionful
parameters in the dark sector Lagrangian at high scales. At low scales, the
Coleman-Weinberg
mechanism~\cite{Coleman:1973jx,Hempfling:1996ht,Chang:2007ki,Khoze:2013uia} then leads to dimensional
transmutation, i.e.~the spontaneous generation of a mass scale through loop effects, which
in turn triggers the dark Higgs mechanism. This set-up enables us to study the profound
connection between DM and GW physics and ask the question: Can the DM mass be generated
through a FOPT that also produces the stochastic GW signal observed by PTAs?

Nearly-conformal dark sectors have been studied in great detail in the context of GWs from
FOPTs~\cite{Baldes:2018emh,Levi:2022bzt, Madge:2023dxc, Sagunski:2023ynd,
  Gouttenoire:2023pxh, Goncalves:2025uwh,
Fujikura:2023lkn,Hosseini:2023qwu,Salvio:2023blb, Ferrante:2023bcz}. It has been shown
that the effective potential changes only slowly with temperature, leading to potentially
strong supercooling and a large energy release during the FOPT. The resulting GW signal
can be correspondingly large and provide a good fit to the observed PTA signal in the
nano-Hertz range. However, previous studies of such dark sectors have typically assumed
that all dark sector states are unstable and decay, such that there is no viable DM
candidate (see however ref.~\cite{Kierkla:2022odc} for a notable exception based on an
$SU(2)$ gauge group or ref.~\cite{Fujikura:2024jto} where asymmetric DM produces the baryon asymmetry in the SM). The model that we consider in the present work can therefore also
be considered as an extension of a nearly-conformal dark sector to include a stable
particle which decouples from thermal equilibrium and plays the role of DM.

Although the dark sector that we consider consists of three new particles, the assumption
of a conformal symmetry strongly reduces the freedom of the model. The dark sector itself
is characterised by only three parameters (the vev $v$, the gauge coupling
$g$, and the Yukawa coupling $y$), out of which two are tightly constrained by the PTA
signal and the third by the relic density requirement. A fourth parameter (the kinetic
mixing $\kappa$) characterises the interactions with SM particles, which determines the
production and decay of the unstable dark sector particles. It is therefore {\it a priori} far from clear
that the model can provide a viable explanation of all available data.

A particular challenge is to reproduce the observed DM relic abundance. It was
shown in ref.~\cite{Bringmann:2023iuz} that the simplest possibility, namely that the DM
particles annihilate directly into dark Higgs bosons, generically implies a dark sector mass scale
around the electroweak scale and thus GW signals in the mHz range potentially detectable with
LISA. When instead considering GWs in the nHz range, it becomes necessary to suppress
the DM annihilation rate in order to avoid depleting the density of DM particles. Here
we consider two possibilities to achieve this goal: forbidden annihilations with a
kinematic threshold~\cite{DAgnolo:2015ujb} and secluded annihilations in a decoupled
dark sector with non-negligible chemical potential~\cite{Pospelov:2007mp}.

We explore in detail the phenomenology of DM from a nearly conformal dark sector and show
that the model can simultaneously fit the PTA signal, reproduce the observed DM relic
abundance and satisfy all other observational constraints. We pay particular attention to
the cosmological evolution of the dark sector after the phase transition, which is
complicated by the fact that both the DM particle and the dark Higgs boson may depart from
thermal equilibrium. The minimal scenario, in which the dark Higgs boson can only decay
via loops involving dark photons, gives tight predictions for the masses and couplings of
all particles, leading to testable predictions for near-future beam-dump experiments. We
also consider an extended scenario, in which higher-dimension operators generate an
additional tree-level coupling to leptons and the constraints can be relaxed.

The remainder of this work is structured as follows. In section~\ref{sec:conformal-models}
we introduce the model that we study and calculate the finite-temperature effective
potential. The two main constraints on the model, namely the GW signal and the DM relic
density, are discussed in sections~\ref{sec:gw} and~\ref{sec:relicdensity}, respectively.
In section~\ref{sec:constraints} we briefly review the various other constraints on the
model. We then present the results from {extensive} global fits of our model
in section~\ref{sec:results}. In section~\ref{sec:outlook}, we discuss our results and the potential impact of future measurements, before concluding in
section~\ref{sec:conclusions}. Additional details on our calculations are provided in the appendices~\ref{app:toy_model}, \ref{app:darkwalls} and \ref{app:freeze-out}.

\section{Conformal Sub-GeV Dark Matter}
\label{sec:conformal-models}

We consider a classically conformal dark sector with a $U(1)^{\prime}$ gauge symmetry. It
comprises a dark Higgs field $\Phi$, the $U(1)'$ gauge boson $A^{\prime}_{\mu}$ (called dark photon)
and two left-handed fermions $\chi_{1,2}$ which have opposite $U(1)'$ charge in order to keep
the model free of anomalies. We set the $U(1)^{\prime}$ charges to $Q_\Phi = 1$,
$Q_1 = - 1/2$, $Q_2 = 1/2$, such that the gauge-invariant Lagrangian is given by
\begin{align}\label{eq:lagrangian-conformal}
   \mathcal{L} = |D_{\mu}\Phi |^2  - \frac{1}{4} F^{\prime}_{\mu\nu}F^{\prime\mu\nu}
  + \bar{\chi}_1 i\slashed{D} \chi_1 + \bar{\chi}_{2} i \slashed{D} \chi_2 - \left(\frac{y_1}{2} \Phi \bar{\chi}_1^c\chi_1 + \frac{y_2}{2} \Phi^* \bar{\chi}_{2}^c\chi_{2} + \text{h.c.}\right)\, - V(\Phi).
\end{align}
The covariant derivative reads $D_\mu = \partial_\mu + i g Q A_\mu^\prime$ and $\chi^c$ denotes the charge
conjugated field. We assume for simplicity that the two fermions are mass-degenerate,
i.e.\ $y_1 = y_2 \equiv y$. This degeneracy could result for example from a symmetry under the
exchange $\chi_1 \leftrightarrow \chi_2^c$, making our assumption technically natural.

The tree-level scalar potential respecting the conformal and $U(1)^{\prime}$ symmetry is given by 
\begin{align}
  V(\Phi) = \lambda (\Phi^{*}\Phi)^2 \,,
\end{align}
where the normalisation of the quartic coupling constant $\lambda$ is conventional.
The conformal symmetry of the classical theory is understood
as the invariance of the action $S = \int \diff^4 x \, \mathcal{L}$ under the set of operations
$x^\mu \rightarrow \alpha x^\mu$, $\Phi(x) \rightarrow \alpha^{-1} \Phi(\alpha x)$,
$A_\mu^\prime(x) \rightarrow \alpha^{-1} A_\mu^\prime(\alpha x)$ and
$\chi(x) \rightarrow \alpha^{-3/2} \chi(\alpha x)$, where
$\alpha > 0$ is a scaling factor. This symmetry also forbids a bare fermion mass term of the form $m_\chi \bar{\chi}^c_1 \chi_2$. The conformal symmetry is explicitly broken by radiative corrections to the
tree-level potential, which introduce a mass term $m^2 \Phi^\ast \Phi$. 

\subsection{Radiative and temperature-dependent corrections}

The breaking of the conformal symmetry can be illustrated by expanding 
$\Phi = (\phi_\text{b} + \phi + \mathrm{i} \varphi)/\sqrt{2}$ around a (homogeneous and static) 
background field value $\phi_\text{b}$.
The leading-order radiative corrections to the potential $V_\text{tree}(\phi_\text{b})= \lambda \phi_\text{b}^4/4$ 
at zero temperature are then given by the usual Coleman-Weinberg potential~\cite{Coleman:1973jx}
\begin{align}\label{eq:coleman-weinberg-potential}
  V_{\mathrm{CW}}(\phi_\text{b}) &= \sum_{a} \eta_{a} g_{a} \frac{m_a^4(\phi_\text{b})}{64 \pi^2} \left[ \log \frac{m_{a}^2(\phi_\text{b})}{\bar{\mu}^{2}} - C_{a} \right]\,,
\end{align}
where the index $a = \bc{\phi, \varphi, A^\prime, \chi_1, \chi_2}$ runs over all fields that obtain masses
for non-zero background field values:\footnote{%
Note that while $\phi$ is a real scalar particle and the dark photon obtains a physical mass due to the spontaneous `breaking' of the $U(1)'$
symmetry for $\phi_\text{b}\neq0$, 
the associated Goldstone mode is {\it not} given by $\varphi$ -- which explains 
the appearance of a mass term in the true vacuum and hence the absence of a shift symmetry. For technical reasons, this specific expansion 
of $\Phi$ is still useful for the underlying computation 
performed in Landau gauge~\cite{Delaunay:2007wb,Andreassen:2014eha}.
}
\begin{align}\label{eq:masses}
  m^2_{\phi}(\phi_\text{b}) = 3\lambda\phi_\text{b}^2\,, && m^2_{\varphi}(\phi_\text{b}) = \lambda\phi_\text{b}^2\,,  && m_{A'}^2(\phi_\text{b}) = g^2\phi_\text{b}^2 \,, 
  && m_{\chi}(\phi_\text{b}) = \frac{y\phi_\text{b}}{\sqrt{2}}\,.
\end{align}
The internal degrees of freedom in eq.~(\ref{eq:coleman-weinberg-potential}) are given by $g_\phi = 1$,
$g_\varphi = 1$, $g_{A^\prime} = 3$ and $g_{\chi_1} =g_{\chi_2} = 2$, and bosons and fermions differ in their
contributions by a sign $\eta_{\phi, \varphi, A^\prime} = + 1$ and
$\eta_{\chi_1, \chi_2} = - 1$. 
We employ the $\overline{\text{MS}}$ scheme, for which the
renormalisation constants read $C_{\phi, \varphi, \chi_1, \chi_2} = \frac{3}{2}$ and
$C_{A^\prime} = \frac{5}{6}$.
Adding to this the tree-level potential and the contributions from counter-terms, with renormalisation 
conditions that ensure $m_\phi(\phi_{\rm b}=0)=0$ and define the renormalised coupling $\lambda$ 
at some energy scale $\Lambda$,
we obtain the zero-temperature effective potential at one-loop as 
\begin{align}
  V_\text{eff}(\phi_\text{b}, T= 0) = \frac{\lambda}{4}\phi^4_\text{b} + \sum_{a} \frac{\eta_ag_a}{64\pi^2} m^4_a(\phi_\text{b}) \left[ \log \ba{\frac{\phi_\text{b}^2}{\Lambda^2}}  - \frac{25}{6}\right]\,.
\end{align}
This potential has a maximum at $\phi_\text{b} = 0$ and a minimum at $\phi_\text{b} = v$ if the
relation
\begin{align}
    \lambda =\sum_a \frac{g_a \eta_a}{48 \pi^2} \frac{m_a^4(v)}{v^4} \bb{11 - 3 \log \ba{\frac{v^2}{\Lambda^2}}} 
    \label{eq:lambdagen}
\end{align}
is satisfied. Since the coupling $\lambda$ is scale-dependent, there will be one specific scale
$\Lambda$ for which eq.~\eqref{eq:lambdagen} is satisfied. Identifying this scale with the vev
of the background field, $\Lambda = \left\langle \phi_\text{b} \right\rangle = v$ yields
\begin{equation}
    \lambda = 11 \sum_a \frac{g_a \eta_a}{48 \pi^2} \frac{m_a^4(v)}{v^4} = \frac{11}{48 \pi^2} \ba{10 \lambda^2 + 3 g^4 - y^4} \, , \label{eq:lambda}
\end{equation}
which determines $\lambda$ as a function of $g$ and $y$. The emergence of a specific energy scale
$v$ from a dimensionless coupling $\lambda$, {ultimately because of a necessarily
  scale-dependent renormalisation condition}, is referred to as \textit{dimensional
  transmutation}. This breaks the conformal symmetry present in the classical theory.
Reinserting the above condition for $\lambda$ into the effective potential, we finally obtain
\begin{align}
    V_\text{eff}(\phi_\text{b}, T = 0) = \sum_a \frac{g_a \eta_a}{64 \pi^2} \frac{m_a^4(v)}{v^4} \phi_\text{b}^4 \bb{\log \ba{\frac{\phi^2_\text{b}}{v^2}} - \frac{1}{2}} \, .
\end{align}
The fact that this potential has a minimum at $\langle \phi_b \rangle = v \neq 0$ implies that 
both the conformal symmetry and
the
$U(1)'$ gauge symmetry are broken at zero temperature, and the various particles obtain
masses proportional to $v$, see eq.~\eqref{eq:masses}.

\begin{figure}[]
  \centering
  \includegraphics[width=0.9\textwidth]{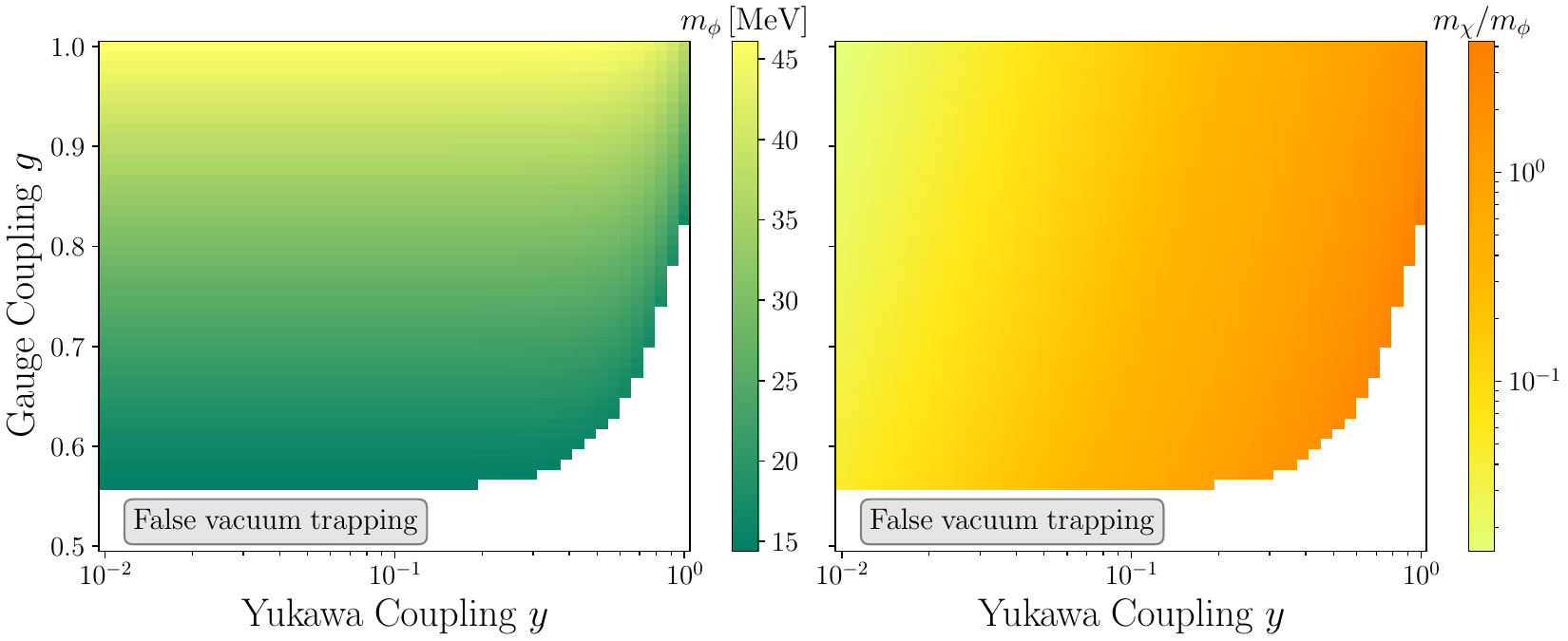}
  \caption{Dark Higgs mass $m_\phi$ and DM mass ratio $m_\chi/ m_\phi$ in dependence of the gauge coupling $g$ and 
  the Yukawa coupling $y$, for a vev $v = 100\,\text{MeV}$.
  }
  \label{fig:masses}
\end{figure}

In figure~\ref{fig:masses} we show, for future reference, how the dark Higgs mass $m_\phi$
and the ratio $m_\chi/m_\phi$ depend on $g$ and $y$. Here, we have fixed $v=100$\,MeV. The main
observation is that $m_\phi$ is independent of $y$ in large parts of the parameter space, so
the ratio $m_\chi / m_\phi$ is determined primarily by the value of $y$. Furthermore, the dark
photon is found to be the heaviest dark sector particle throughout the relevant parameter
space. These features will help us to interpret the results of our global analyses in
section~\ref{sec:results}. For too small gauge couplings
$g \lesssim 0.55$, the phase transition cannot finish regardless of the specific
value of $y$. This is due to the potential barrier persisting for too long, resulting in
the background field being trapped in the false vacuum (see section~\ref{sec:gw}).

At finite temperatures the effective potential receives additional corrections, given at leading order by
\begin{equation}
  V_T(\phi_\text{b}) = \frac{T^4}{2\pi^2} \sum_a \eta_a g_a J_{\text{b}/\text{f}}\left(\frac{m_a^2(\phi_\text{b})}{T^2}\right)\,,
  \label{finiteT}
\end{equation}
where the thermal functions $J_{\text{b}/\text{f}}$ for bosons/fermions are defined in
ref.~\cite{Quiros:1999jp}. The finite-temperature potential in eq.~\eqref{finiteT} suffers
from infrared divergences for bosonic modes when $T \gg m_a$, leading to a breakdown of
finite-temperature perturbation theory~\cite{Athron:2023xlk}. The resummation of these
modes is commonly done by ``daisy resummation''~\cite{Carrington:1991hz}. Here we follow
the Arnold-Espinosa prescription~\cite{Arnold:1992rz}, thus only resumming the
Matsubara zero-modes, which effectively leads to an additional contribution to the
effective potential of the form
\begin{equation}
  V_{\rm daisy}(\phi_\text{b}) = -\frac{T}{12\pi}\sum_{a=\phi,\varphi,A_\text{L}^\prime} g_a \left[(m_a^2(\phi_\text{b}) + \Pi_a(T))^{3/2} - (m_a^2(\phi_\text{b}))^{3/2}\right],
\end{equation}
where $a$ now only runs over the scalar fields $\phi$ and $\varphi$ and the longitudinal component of the dark
photon $A_\text{L}^\prime$. The one-loop thermal masses $\Pi_a$ in our model read
\begin{align}
  \Pi_\phi = \left(\frac{\lambda}{3} + \frac{y^2}{12} + \frac{g^2}{4}\right)T^2\,, &&
  \Pi_\varphi = \left(\frac{\lambda}{3} + \frac{y^2}{12}+\frac{g^2}{4}\right)T^2\, && \text{and} &&
  \Pi_{A^\prime} = \frac{5}{12} g^2T^2 \,.
\end{align}

\begin{figure}[t]
  \centering
  \includegraphics[width=.6\textwidth]{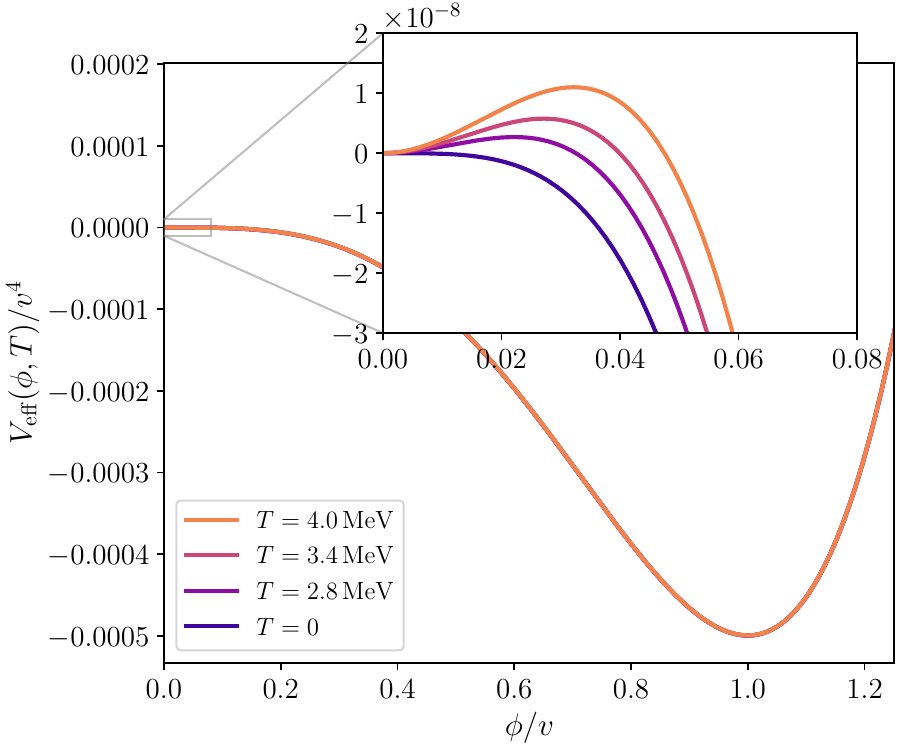}
  \caption{Effective potential and zoom-in on the potential barrier for the best-fit point
    obtained for the \textit{coupled dark sector} scenario with $v = 173\,$MeV (see
    table~\ref{tab:bestfit}). For this parameter point the nucleation of bubbles happens
    at $T_\text{nuc} = 2.8\,$MeV. The energy released during the FOPT then reheats
    the universe to $T_\text{reh} = 16.8\, \mathrm{MeV}$.}
  \label{fig:potential-plots}
\end{figure}

In total, the effective potential is thus given by
\begin{equation}
 V_{\rm eff}(\phi_\text{b}, T) = V_\text{eff}(\phi_\text{b}, T = 0)
 + V_T(\phi_\text{b}, T) 
 + V_{\rm daisy}(\phi_\text{b}, T)\, .\label{eq:effectivepotential}
\end{equation}
Since the thermal part of the potential induces terms of the form $T^2 \phi_\text{b}^2$, the
$U(1)'$ symmetry is restored at high temperatures. 
Further, transitions between the $\phi_\text{b} = 0$ and $\phi_\text{b} = v$ phase are possible.
Interestingly, these transitions can be strongly supercooled due to the weak, logarithmic
temperature dependence of the height of the potential barrier. This feature makes
classically conformal dark sectors particularly interesting in the context of GW signals
from strong FOPTs.
In figure~\ref{fig:potential-plots} we show the effective potential $V_{\rm eff}(\phi_\text{b}, T)$ for the best-fit point of the 
coupled dark sector scenario discussed in detail in section~\ref{sec:results}. 
The weak temperature dependence of the potential barrier just mentioned is clearly visible. 
The barrier hence persists for a large range of
temperatures, leading to strong supercooling and eventually strong GW
signals.

\subsection{Interaction terms}
\label{sec:SM_interactions}

After spontaneous symmetry breaking, the dark Higgs boson generates mass terms for the two Majorana fermions $\chi'_{1,2} = \chi_{1,2} + \chi_{1,2}^c$ with the following interaction Lagrangian:
\begin{align}
  \mathcal{L}_{\mathrm{int}}
  \supset - \lambda v \phi^3  - \frac{\lambda}{4}\phi^{4}+ g^2v A'_{\mu}A'^{\mu}\phi + \frac{g^2}{2} \phi^2 A'_{\mu}A'^{\mu} 
    + \sum_i \frac{g}{2} Q_i \bar{\chi'_i} \gamma^\mu \gamma^{5} \chi'_i A^\prime_\mu - \frac{y}{2\sqrt{2}}\phi \bar{\chi'_i}\chi'_i \,. 
\end{align}
Since the two fermions are assumed to be mass-degenerate, their mass terms and the interactions with the dark Higgs boson can be written in a more economic form by defining a single Dirac
fermion~\cite{Choi:2008pi}
\begin{equation}
    \chi = \frac{1}{\sqrt{2}} \left(\chi_1' + i \chi_2'\right) \; ,
\end{equation}
such that $\mathcal{L} \supset -m_\chi \bar{\chi} \chi - \tfrac{y}{\sqrt{2}} \phi \bar{\chi} \chi$. The interactions with the dark 
photon cannot be simplified in this way, since the two Majorana fermions couple to the $A'$ with opposite sign. 
The sign of $Q_i=\pm1/2$, however, does not enter
in any of the processes that we consider in the remainder of this work.\footnote{
Note that such sign differences can only impact the interference terms between diagrams
involving both $\chi_1$ and $\chi_2$. At tree-level, this would for example be the case for $A'$ and $\phi$ mediated 
$\chi_1$-$\chi_2$ scattering, respectively, but in our calculations we anyway neglect scattering through $A'$ exchange 
because  $A'$ is much heavier than $\phi$.
}
For these processes, we can therefore also simplify the interaction with the dark photon by introducing
an effective interaction term $\mathcal{L} \supset \tfrac{g}{2} \bar{\chi} \gamma^\mu \gamma^5 \chi A'_\mu$. We will use this form in our subsequent calculations.

In principle, the dark sector introduced above does not require interactions with the SM
for the spontaneous breaking of the $U(1)'$ symmetry to occur. However, if the dark sector
contributes significantly to the energy density of the universe at high temperatures, its
energy density must eventually be transferred to the SM plasma in order to recover the
known cosmological evolution at low temperatures~\cite{Bringmann:2023opz}. In particular, the various massive
particles arising in the dark sector after symmetry breaking must annihilate or decay
efficiently enough into SM particles in order to not overclose the universe.

Thus, a portal coupling of the dark sector to the SM is required. The dark sector that we
have introduced in general permits for two such portals at the renormalisable level.
First, the dark Higgs field can couple to the SM Higgs field through a term
$\lambda_{h \phi} H^2 \Phi^2$ in the scalar potential. This term would however spoil the classically
conformal symmetry of the dark sector. In this work we therefore focus on the alternative
possibility that the dark photon couples to the photon $A_{\mu}$ via kinetic
mixing: \begin{align}\label{eq:kinetic-mixing} \mathcal{L} \supset \frac{\kappa}{2} F_{\mu\nu} F^{\prime\mu\nu} \,,
\end{align}
where $F'_{\mu\nu}$ and $F_{\mu\nu}$ denote the respective field strength tensors and $\kappa$ is the kinetic mixing 
parameter. Crucially, the kinetic mixing portal does not violate the conformal invariance 
required to obtain strong supercooling.

If the dark photon mass is sufficiently light compared to the $Z$ boson, we can neglect
the latter and obtain an effective interaction between $A_{\mu}^{\prime}$ and the SM fermions:
\begin{align}\label{eq:effective-mixing}
  \mathcal{L}_{\mathrm{int}} = - \kappa e A^{\prime}_{\mu} \sum_f q_f \bar{f} \gamma^{\mu} f\,,
\end{align}
where the sum is over all SM fermions and $q_{f}$ denotes the respective electric charges. 

For dark photon masses above the MeV scale, kinetic mixing allows for efficient decays of
the dark photon into electron-positron pairs, as well as efficient scattering of dark
sector particles off electrons and positrons in the plasma. In the parameter range of
interest ($10^{-7} < \kappa < 10^{-3}$), we can therefore assume that the dark sector
shares a common temperature with the SM thermal bath before DM freeze-out. The FOPT in the dark sector therefore does not only reheat the dark sector but also the SM bath, in contrast to the discussion in ref.~\cite{Bringmann:2023opz}.
The dark fermion $\chi$ remains stable also in the presence of kinetic mixing and therefore
constitutes a potential DM candidate. In particular, it can annihilate into SM fermions via an off-shell dark photon, and we 
will study the resulting relic abundance in detail in section~\ref{sec:relicdensity}. 
  
On the other hand, kinetic mixing allows for decays of the dark Higgs bosons into SM fermions via a loop involving two dark photons~\cite{Batell:2009yf}. For
heavy dark photons $m_{A'} \gg m_\phi$, the corresponding decay width is given by
\begin{equation}
\label{eq:dark_higgs_decay}
    \Gamma_{\phi \to f\bar{f}} = \frac{27 \, e^4 \kappa^4 g^2 m_\phi}{128 \pi^5} \frac{m_f^2}{m_{A^\prime}^2} \, 
\end{equation}
for a SM fermion with mass $m_f \ll m_\phi$. If only decays into electrons are kinematically allowed, one obtains approximately
\begin{equation}
    \label{eq:tauphi}
    \tau_\phi \approx 2500 \, \mathrm{s} \left(\frac{\kappa}{10^{-4}}\right)^4 \left(\frac{g}{0.75}\right)^2 \frac{m_\phi}{30 \, \mathrm{MeV}} \left(\frac{m_{A^\prime}}{100 \, \mathrm{MeV}}\right)^{-2} \, . 
\end{equation}
Hence the loop-induced Higgs decays are slow compared to the expansion of the universe
before Big Bang Nucleosynthesis (BBN) and therefore not sufficient to equilibrate the dark
and visible sectors at early times. The same is true for all other processes that do not
involve the fermionic DM particle $\chi$, such as $\phi \to 4 e$ or $\phi + e \to 3 e$. In contrast to
some recent claims in the literature~\cite{Costa:2025csj}, we find that kinetic mixing
alone is not sufficient to fully deplete the energy density of the dark sector between the
FOPT (which typically happens around $1$\,MeV, see
figure~\ref{fig:potential-plots} and section~\ref{sec:gw}) and BBN.

In the following, we will therefore also consider the possibility that additional
interactions are present that allow the dark Higgs boson to decay quickly into SM
particles. The simplest option would be to induce such decays via mixing of the dark Higgs
boson with the SM Higgs boson after spontaneous symmetry breaking. Such a mixing would
however also induce a dark Higgs boson mass after electroweak symmetry breaking, thus
spoiling the conformal symmetry of the model~\cite{Madge:2023dxc}. A simple alternative
would be to consider an effective coupling to photons, which is however tightly
constrained by the combination of experimental data and the observation of
SN1987A~\cite{Balazs:2022tjl,Antel:2023hkf} and not viable in the mass range of interest.
Another possibility of increasing the dark Higgs decay width is to assume an extended
electroweak Higgs sector coupled to the dark Higgs through mass mixing~\cite{Han:2023olf}.
Here we instead consider an effective coupling of the dark Higgs bosons to electrons,
$y_\text{eff} \phi \bar{e} e$, for which supernova constraints are
weaker~\cite{Ferreira:2022xlw,Hardy:2024gwy}. A specific example for how such an
interaction may be generated is presented in appendix~\ref{app:toy_model}. While such a
higher-dimensional operator violates scale-invariance, its effect on the dynamics of the
phase transition is strongly suppressed through the high scale of new physics responsible
for this operator.

\begin{figure}[t]
    \centering
    \includegraphics[width=0.49\linewidth]{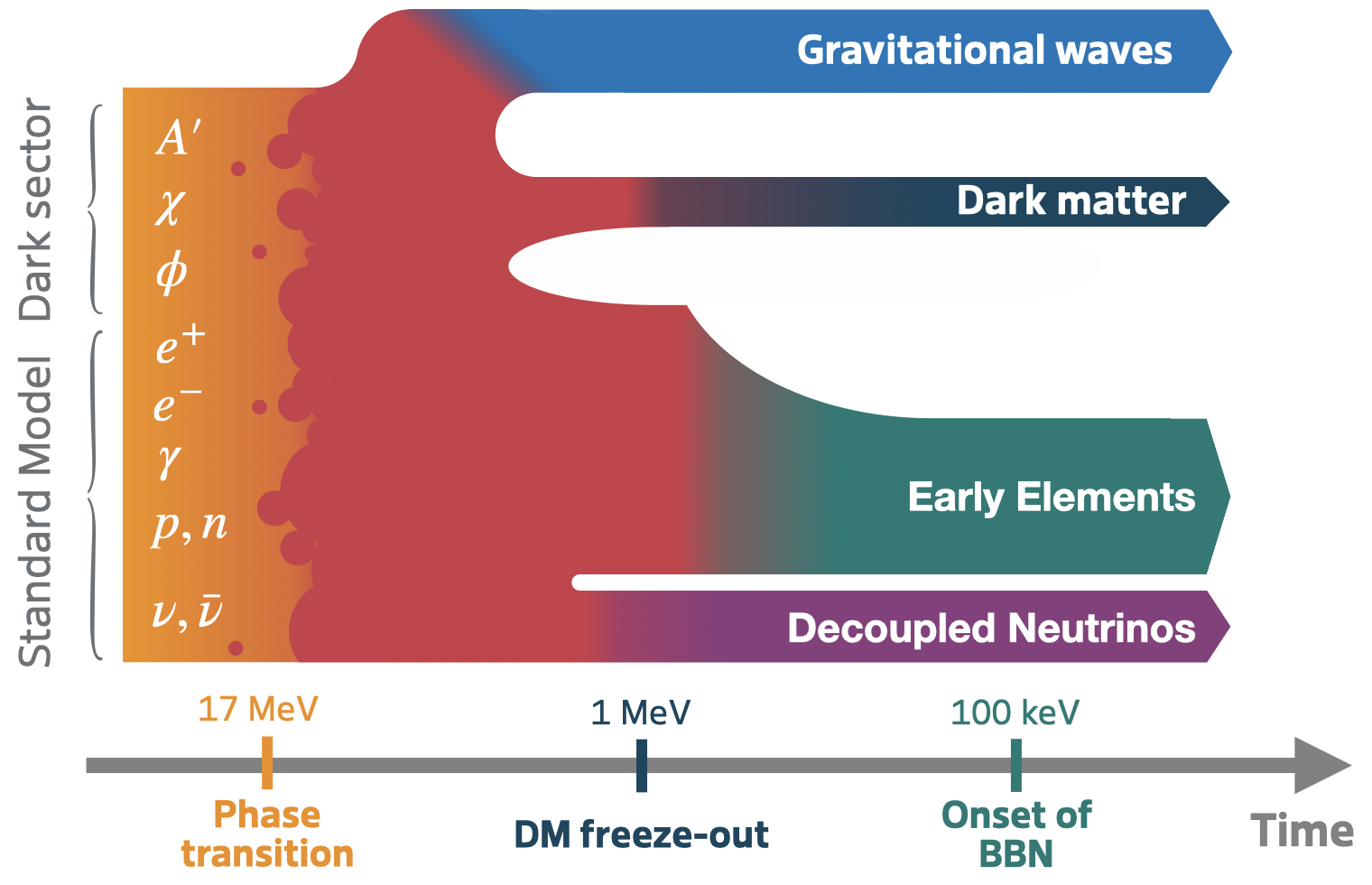}\hfill
    \includegraphics[width=0.49\linewidth]{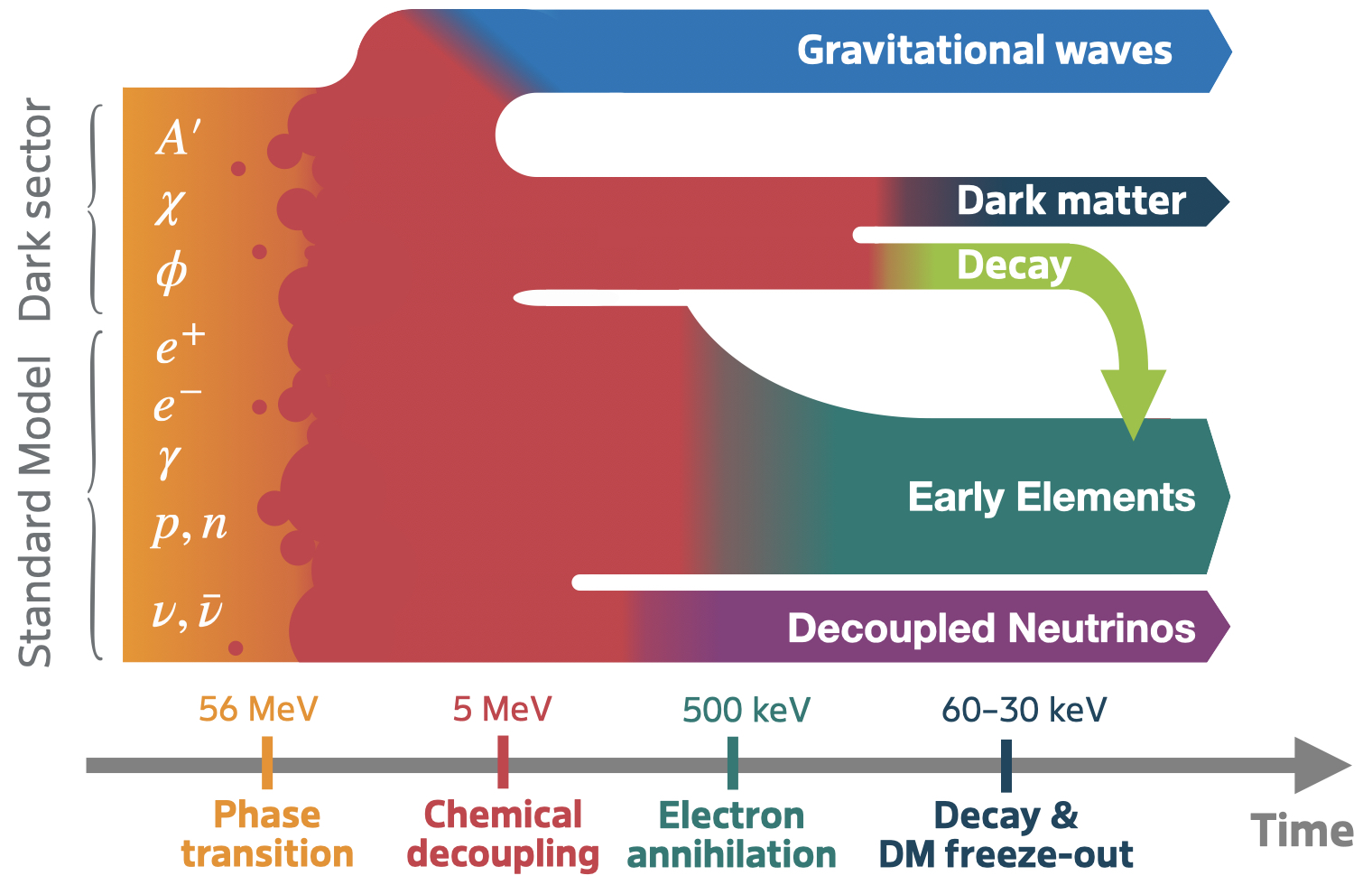}
    \caption{Cosmological history of the scenario with a coupled dark sector (left) and a
      secluded dark sector (right). The main difference between them is that in the former
      case the dark Higgs bosons are always in equilibrium with the SM thermal bath,
      whereas in the latter case they decouple and decay during or after BBN. See
      table~\ref{tab:bestfit} for typical parameter points for these scenarios.}
    \label{fig:history}
\end{figure}

In the following, we will refer to the case of a dark Higgs boson with such an effective
coupling to electrons as \emph{coupled dark sector} and the case where the dark Higgs
boson can decay only through dark photon loops as \emph{secluded dark sector}.\footnote{We
  emphasise that our use of the term ``secluded'' differs somewhat from the one in
  ref.~\cite{Li:2025rrs}, which considers the case that the dark sector is not in thermal
  contact with the SM during the FOPT. In our set-up, the two sector are in thermal
  contact during the phase transition, but decouple subsequently.} The different
cosmological history of these two cases is illustrated in figure~\ref{fig:history}. We
emphasise that in the secluded dark sector case, a non-negligible abundance of dark Higgs
bosons remains after BBN, leading to additional cosmological constraints. The impact of
this assumption on the relic density calculation is discussed in
section~\ref{sec:relicdensity}, while section~\ref{sec:constraints} considers other
relevant constraints on the two scenarios. The impact of these constraints on the allowed
parameter space will be the topic of section~\ref{sec:results}.

In summary, the dark sector that we have introduced is described by three independent
parameters, namely the gauge coupling $g$, the Yukawa coupling $y$ and the vev $v$. The
quartic coupling $\lambda$ can be calculated from these parameters following
eq.~\eqref{eq:lambda}. The dark sector contains three new particles: a dark Higgs boson
$\phi$, a dark photon $A'$ and a Dirac fermion DM candidate $\chi$, with masses as given in
eq.~\eqref{eq:masses}. Finally, we have introduced the kinetic mixing parameter $\kappa$ as a
fourth independent parameter of our model.

For the following discussion, it will be convenient to consider two specific parameter points, which are summarised in 
table~\ref{tab:bestfit}. The first corresponds to the best-fit point of the coupled dark sector scan presented in 
section~\ref{sec:results_coupled}. The second corresponds to the best-fit point of the secluded dark sector scan after 
imposing an additional requirement on the lifetime of the dark Higgs boson, see section~\ref{sec:results_secluded}. We will 
refer to these two points as Point A and Point B in the following.

\section{The GW signal detected by PTAs}
\label{sec:gw}

Following the announcement of evidence for a GW background at nHz
frequencies~\cite{NANOGrav:2023gor, Reardon:2023gzh, EPTA:2023fyk, Xu:2023wog,
  Miles:2024seg}, various possible sources for this novel signal have been explored, the
most prominent one being SMBH binaries~\cite{NANOGrav:2023hfp, NANOGrav:2023pdq}. Even
though GW signals from SMBH binaries have been a subject of research for the past
decades~\cite{Rajagopal:1994zj, Jaffe:2002rt, Wyithe:2002ep, Sesana:2004sp,
  McWilliams:2012an, Burke-Spolaor:2018bvk, Simon:2023dyi}, the predictions are still
troubled by astrophysical uncertainties. While it appears rather certain that there is
\textit{some} contribution to the observed signal from SMBH binaries, it is far from clear
whether this contribution is dominant~\cite{NANOGrav:2023hvm, Chen:2025wel}. Arguably more
exciting, the dominant source of the observed GW signal could then also be of cosmological
origin, such as a FOPT at temperatures of around $100 \, \text{MeV}$. In this work we
therefore allow for both options, leading to a combined signal of the form
\begin{align}
    h^2 \Omega_\text{gw}(f) = h^2 \Omega_\text{gw}^\text{BH}(f) + h^2 \Omega_\text{gw}^\text{PT}(f) \, .
\end{align}
In the following two subsections we will briefly review
the predictions for these two contributions, before discussing the PTA likelihood employed in our global fit
in section \ref{sec:PTAlike}.

\begin{table}[t]
    \centering
    \resizebox{\textwidth}{!}{%
    \begin{tabular}{r|ccccc|ccc|cccc}
        \toprule
         Parameter & $g$ & $y$ &  $\delta$ & $\kappa$ & $v$ & $m_{A^\prime}$ & $m_\phi$ & $m_\chi$  & $T_\text{p}$ & $T_\text{reh}$ & $\alpha$ & $\beta/H$ \\
         Unit & - & - & - & - & MeV & MeV & MeV & MeV & MeV& MeV & - & - \\
        \midrule
        Point A & 0.677 & 0.224 & $-0.322$ & $ 2.29\times10^{-6}$ & 173 & 117 & 36.3 & 27.4 & 2.28 & 16.8 & $4.7\times10^{3}$ & 33.7 \\
       Point B  & 0.601 & 0.234 & $6.48 \times 10^{-3}$ & $2.7\times10^{-4}$ & 692 & 416 & 114 & 115 & 1.21 & 55.9 & $9.48\times10^{6}$ & 18.0 \\
        \bottomrule
    \end{tabular}%
    }
    \caption{Particle masses and phase transition properties for two specific parameter
      points. The first corresponds to the best-fit point of the coupled dark sector scan
      presented in section~\ref{sec:results_coupled}. The second corresponds to the
      best-fit point of the secluded dark sector scan after imposing an additional
      requirement on the lifetime of the dark Higgs boson, see
      section~\ref{sec:results_secluded}. The derived parameter
      $\delta = (m_\chi - m_\phi)/m_\chi = 1 - \sqrt{6 \lambda}/y$ quantifies the relative mass difference
      between the DM particle and the dark Higgs boson.}
    \label{tab:bestfit}
\end{table}

\subsection{GW contribution from supermassive black hole binaries}
\label{sec:GWSMBHB}

The GW signal emitted by a population of inspiraling SMBHs can be calculated by
integrating the rate of SMBH merger events multiplied with the GW signal of a given merger
over the cosmic evolution~\cite{Phinney:2001di, Wyithe:2002ep}. The SMBH merger rate
depends on the galaxy merger rate, the mass distribution of the SMBH binaries and details
of the binary evolution. The latter depends on how the initial binding energy of the dual
system is dissipated on a sub-Hubble timescale. The mechanism of stellar scattering
remains very difficult to model, and it is not yet understood if this mechanism can give
the SMBH merger rate required to fit the PTA data. This is the so-called final parsec
problem~\cite{Binney:1987, Milosavljevic:2002ht,
  Dosopoulou:2016hbg,Alonso-Alvarez:2024gdz}. Depending on the astrophysical modeling of
the SMBH binary populations and their merger history, a wide range of signal predictions
have been obtained in the past. All of these, however, roughly follow a power-law shape of
the form
\begin{align}
    h^2 \Omega_\text{gw}^\text{BH}(f) = \frac{2 \pi^2}{3 H_{100}^2} A^2 \ba{\frac{f}{1 \, \text{yr}^{-1}}}^{5-\gamma} \, \text{yr}^{-2} \, ,
\end{align}
where $H_{100} \equiv H_0 / h = 100 \, \text{km} \, \text{s}^{-1} \, \text{Mpc}^{-1}$
and $H_0 = 67.8 \, \text{km} \, \text{s}^{-1} \, \text{Mpc}^{-1}$ is the present-day Hubble rate. 
The amplitudes $A$ range between 
$10^{-18}$ and $10^{-14}$ and slopes $\gamma$ between 3 and 6, see ref.~\cite{NANOGrav:2023hfp} and references in 
their table A1. 
We will assume that the SMBH binary contribution follows the
above power-law, and treat the parameters $A$ and $\gamma$ as nuisance parameters constrained by a
Gaussian likelihood with
\begin{align} \label{eq:prior-smbh}
    \mu_{(A, \gamma)} = \ba{\begin{matrix}
        -15.615\\ 4.707
    \end{matrix}} && \text{and} && \text{cov}_{(A, \gamma)} = \ba{\begin{matrix}
        0.279 \qquad -0.003\\
        -0.003 \qquad 0.124
    \end{matrix}} \, ,
\end{align}
as recommended by the NANOGrav collaboration~\cite{NANOGrav:2023hvm} and
used in their PTArcade~\cite{Mitridate:2023oar} tool.

\subsection{GW contribution from a supercooled phase transition}
\label{sec:GWFOPT}

The GW signal from a supercooled FOPT depends on a number of relevant ingredients and parameters. 

\paragraph{Percolation} 
We trace the minima of the effective potential in eq.~\eqref{eq:effectivepotential} 
as a function of temperature to identify possible vacuum transitions and then calculate
the respective bubble nucleation rates. To do so, we compute the Euclidean bounce action
\begin{align}
 S_3(T) = \int \diff^3 x \bb{\frac{\ba{\nabla \bar{\phi}_\text{b}}^2}{2} + V_\text{eff}(\bar{\phi}_\text{b}, T)} \, ,
\end{align}
where $\bar{\phi}_\text{b}$ is the bounce solution of the $O(3)$-symmetric Klein-Gordon equation. The bubble nucleation rate per unit time and volume is then
determined by~\cite{Linde:1981zj}
\begin{align}
    \Gamma(T) = A(T) \exp \bb{- \frac{S_3(T)}{T}} \, , 
\end{align}
where we approximate $A(T) \simeq T^4$ (see ref.~\cite{Gould:2021dzl} for further discussion). 
We then compute the false vacuum fraction
\begin{align} \label{eq:Pf}
    P_\text{f}(T) = \exp \bb{ -\frac{4 \pi}{3} \int^{T_c}_T \frac{\diff{T^\prime} \, \Gamma(T^\prime) }{H(T^\prime) {T^\prime}^4} \, \bb{\int^{T^\prime}_T \frac{\diff{\tilde{T}}}{H(\tilde{T})} 
    }^3} \, .
\end{align}
Here $T_\text{c}$ denotes the critical temperature, i.e.~the point 
at which the minima of the potential become degenerate,
and the Hubble parameter $H(T)$ includes the contribution of 
the vacuum energy to the total energy density. 
In arriving at eq.~\eqref{eq:Pf} we used $\dot{T} \simeq - H(T) T$,  
see  ref.~\cite{Athron:2023xlk} for a more detailed discussion, and assumed that the reheating of the dark sector within the 
bubbles is instantaneous and that
the bubble wall velocity is $v_\text{w} = 1$ at all times. The assumption of
instantaneous reheating was explicitly checked in ref.~\cite{Bringmann:2023iuz} in a similar model setup, and shown to hold
due to the large intra-dark sector couplings. 
In appendix \ref{app:darkwalls} we discuss in detail why the assumption of large bubble wall
speeds holds throughout our model parameter space. 
Finally, we numerically solve $P_\text{f}(T_\text{p}) = 0.71$ for the percolation temperature $T_\text{p}$ \cite{Ellis:2018mja},
which we use to compute a number of thermodynamic quantities relevant for the prediction of the GW signal. We label the parameter region where no solution can be found as ``False vacuum trapping''.

\paragraph{The mean bubble separation} The first such quantity is the characteristic length scale~\cite{Athron:2023xlk}
\begin{align}
    \label{eq:RH}
  R H_* \equiv \frac{H_\text{p}}{\bb{n(T_\text{p})}^{1/3}}
  && \text{with} && n(T) = T^3 \int^{T_c}_T \frac{\diff{T^\prime}}{H(T^\prime) {T^\prime}^4} \, \Gamma(T^\prime) P_\text{f}(T^\prime) \,,
\end{align}
determined by describing the mean bubble separation at percolation.
The characteristic inverse time scale of the phase transition, i.e.\ the phase transition speed, is often estimated from the 
temperature dependence of the bounce action, 
$\left.\mathrm{d}(S_3/T) / \mathrm{d} \log T\right|_{T_\text{p
}}$, which however does not give a good approximation in the case of strong supercooling, because $\log \Gamma$ no 
longer grows linearly with time~\cite{Athron:2022mmm}.
We therefore determine the phase
transition speed directly from the mean bubble separation and define
\begin{align} \label{eq:mapbetaR}
    \beta/H = \frac{(8 \pi)^{1/3}v_\text{w}}{R H_*}  \, .
\end{align}
The length scale $RH_*$ also directly links to the
quantities used in numerical simulations~\cite{Hindmarsh:2013xza}. We therefore compute
the mean bubble separation directly using eq.~\eqref{eq:RH} but express our results in terms of 
the quantity $\beta/H$ in order to allow a direct comparison with the existing literature on phase transitions. As the 
temperature dependence of the bounce action is only very mild, $\beta/H$ is very small throughout the studied parameter 
space.

\paragraph{Energy budgets} In order to compute the kinetic energy fraction $K$ setting the GW spectral amplitude, we 
employ the prescription developed
in ref.~\cite{Giese:2020znk} and use the pseudo-trace of the energy-momentum tensor of the effective potential 
$V_i(T) \equiv V(\phi_i, T)$ in both the true vacuum ($i = \text{t}$) and the false vacuum ($i = \text{f}$)
\begin{align}
  \bar{\theta}_i = \frac{1}{4} \bb{\rho_i(T) - \frac{p_i(T)}{c_{\text{s}, i}^2(T)}} \simeq V_i(T) - \frac{1}{4} T\partial_T V_i(T)
  && \text{for} && c_{\text{s},i}^2(T) = \frac{\partial_T V_i(T)}{T \partial_T^2 V_i(T)} \approx \frac{1}{3} \, .
\end{align}
Here we used $\rho_i(T) = V_i(T) - T \partial_T V_i(T)$ and $p_i(T) = - V_i(T)$. We have checked
explicitly that the approximation $c_\text{s}^2 \approx 1/3$ holds up to $\mathcal{O}(1\, \%)$ within our
allowed parameter regions in both the symmetric and broken phase.
The kinetic energy fraction is then given by
\begin{align}
    K = 0.6\, \kappa(\alpha) \frac{\alpha}{\alpha + 1} && \text{with} && \alpha = \frac{\bar{\theta}_\text{f}(T_\text{p}) - \bar{\theta}_\text{t}(T_\text{p})}{\rho_\text{rad}(T_\text{p}) } \, ,
\end{align}
where $\rho_\text{rad}$ denotes the total energy density in radiation (SM and dark sector).
For the efficiency factor $\kappa(\alpha)$ we employ the fitting formula in the limit
$v_\text{w} \rightarrow 1$ from the appendix of reference~\cite{Espinosa:2010hh}. Note that the
factor $0.6$ in the kinetic energy fraction accounts for the efficiency of producing
kinetic energy in the bulk fluid motion with respect to the single bubble case, as found
in ref.~\cite{Jinno:2022mie}. Due to the high amount of supercooling, we find that
$\alpha \gg 1$, $\kappa \rightarrow1$ and $K \rightarrow 1$ in the parts of model parameter space relevant in this work,
thus saturating to the strongest GW amplitudes that can possibly be reached by a FOPT.

\begin{figure}[t]
  \centering
  \includegraphics[width=1.\textwidth]{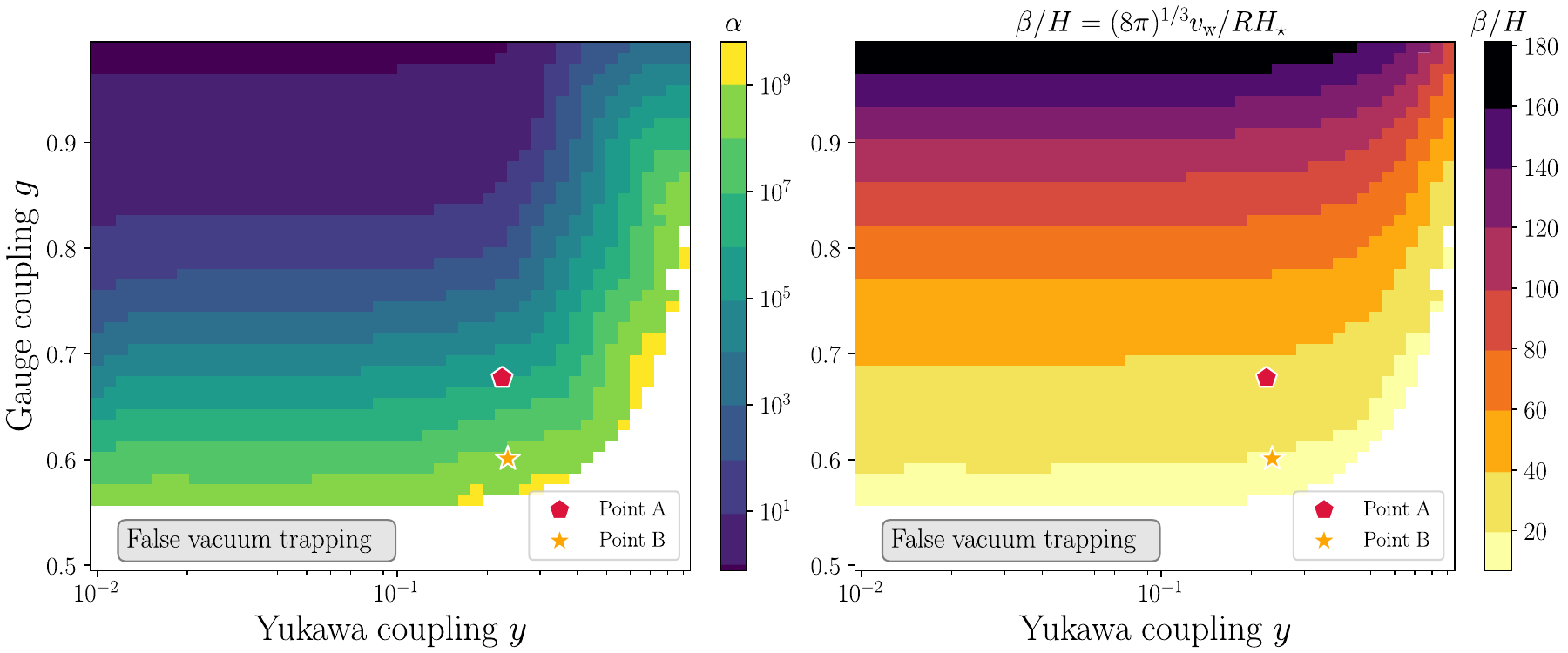}
  \caption{ GW parameters $\alpha$ and $\beta / H$ as a function of $g$ and $y$. The coupling values of the parameter points A 
  and B from table~\ref{tab:bestfit} are indicated by the red pentagon and the yellow star, respectively. For concreteness, we 
  have set $v = 100\,$MeV, but the results are almost completely independent of the vev.}
  \label{fig:pt-alpha-betaH}
\end{figure}

Figure~\ref{fig:pt-alpha-betaH} shows the dependence of $\alpha$ and $\beta/H$ on the model parameters $y$ and $g$ for 
a fixed vev $v = 100\,\text{MeV}$. As the kinetic energy fraction $K \simeq 1$ is almost independent of $\alpha$ for 
$\alpha \gg 1$, the region of parameter space with minimal $\beta/H$ (i.e., the largest bubbles at percolation) leads to the 
strongest GW signals.
The vev $v$ is the only dimensionful quantity in the computation of $\alpha$ and $\beta/H$, neglecting a weak dependence 
on the number of degrees of freedom of the plasma. The parameters $\alpha$ and $\beta/H$ are hence
almost independent of $v$, whereas $v$ sets the scale of the peak frequency,
$f_\text{p}^0 \propto v$. A naive dimensional analysis based on eqs.~(\ref{eq:fbreaks},\,\ref{eq:Hredshift}) and
realising that PTAs probe nano-Hertz frequencies already indicates, that $v \simeq \mathcal{O}(100 \, \text{MeV})$
is required to account for the observed signal.

\paragraph{GW signal} As discussed in appendix~\ref{app:darkwalls}, the bubble walls will quickly reach a terminal, yet 
highly relativistic, velocity. For that reason, the sound wave contribution to the GW spectrum is expected to strongly 
dominate over the colliding bubble
wall contributions (scaling with $1 - \kappa$)~\cite{Caprini:2015zlo}. To determine the timescale on which sound waves
contribute to the GW signal, we introduce the factor
\begin{align}
    \mathcal{Y}_\text{sw} = \text{min}\ba{1, \tau_\text{sh} H_\text{p}} \simeq \text{min}\ba{1, \sqrt{\frac{\Gamma}{K}} R H_*} \simeq \text{min} \ba{1, \frac{2}{\sqrt{3}}R H_* } \, ,
\end{align}
where we assumed that this timescale is determined by the shock formation time, dependent on the adiabatic index 
$\Gamma = 4/3$ and the kinetic energy fraction $K \approx 1$. In addition to sound waves, also turbulence driven by bulk 
fluid motion in the plasma could potentially contribute to the GW spectrum.
The effect of turbulence in cosmological phase transitions is
still not well understood. 
We neglect it in our analysis, assuming that it gives a subdominant contribution 
that only changes the slope of the spectrum away from the 
peak~\cite{RoperPol:2019wvy, Caprini:2018mtu}.

We can hence use the template advertised in ref.~\cite{Caprini:2024hue}, based on the
simulations presented in ref.~\cite{Jinno:2022mie,Caprini:2024gyk}, which we parameterise
as
\begin{align}
    \label{eq:GWspec}
    h^2 \Omega_\text{gw}^\text{PT}(f) = \mathcal{R}h^2 A_{\mathrm{sw}} K^2 \mathcal{Y}_{\mathrm{sw}} \ba{RH_*}  \tilde{S}(f)\,,
\end{align}
where the coefficient $A_{\mathrm{sw}} \approx 0.11$ is determined by simulations. The spectral
shape describes a double broken power law
\begin{align}\label{eq:spectralShape}
  S(f) &= N \left( \frac{f}{f_1} \right)^{3} \left[ 1 + \left( \frac{f}{f_1} \right)^2 \right]^{-1}
  \left[ 1 + \left( \frac{f}{f_2} \right)^4 \right]^{-1}\,,
\end{align}
with the normalisation
\begin{align}\label{eq:stilde}
  \tilde{S}(f) &= \frac{1}{\pi} \left( \sqrt{2} +   \frac{2\,f_2/f_{1}}{1 + f_2^2/f_1^2}  \right) \frac{S(f)}{S(f_{2})}\; ,
\end{align}
and the two frequencies breaks
\begin{align}\label{eq:fbreaks}
  f_1 &\simeq  0.2 \ba{\frac{H_{*,0}}{RH_*} }\,, ~~~~ f_2 \simeq 0.5\, \Delta_{w}^{-1} \ba{\frac{H_{*,0}}{RH_*} }\;.
\end{align}
The dimensionless factor
$\Delta_{w} = \left| v_{w} - c_{s} \right|/\max(v_{w}, c_{s})$ parametrizes the sound shell
thickness and the redshift factor is given by
\begin{align}\label{eq:Hredshift}
  H_{*, 0} &=  \, 16.5 \, \text{nHz} \,\ba{\frac{T_\text{reh}}{100 \, \text{MeV}}}  \ba{\frac{g_{\text{reh}}}{100} }^{1/2}
             \ba{\frac{100}{h_{\text{reh}}}}^{1/3}\,.
\end{align}
The redshift of the energy density given by
\begin{align}
  \mathcal{R}h^2 = \ba{\frac{a_\text{reh}}{a_0}}^4 \ba{\frac{H_\text{reh}}{H_0}}^2 h^2
  = \Omega_{\gamma}h^2 \ba{\frac{h_{0}}{h_{\text{reh}}}}^{4/3} \frac{g_\text{reh}}{g_\gamma} \, ,
\end{align}
where $\Omega_\gamma h^2 = 2.473 \cdot 10^{-5}$ is the present radiation energy
density~\cite{Planck:2018vyg}. The quantities $g_\text{reh}$ and $h_\text{reh}$ refer to
the combined SM and DS effective energy and entropy degrees of freedom at reheating, while
$g_\gamma = 2$ and $h_{0} = 3.91$. All computations presented in this subsection, from the
computation of the effective potential to the GW signal
$h^2 \Omega_\text{gw}^\text{PT}(f)$, are performed using the publicly available code
\textsf{TransitionListener}~\cite{Ertas:2021xeh} which we adapted to use $RH_*$ to
estimate the characteristic GW frequency.

\paragraph{Reheating temperature} The relevant scale for setting the characteristic
frequency of the signal in eq.~\eqref{eq:fbreaks} is the reheating temperature
$T_\text{reh} \neq T_\perc$, Assuming instantaneous reheating inside the bubbles and
employing energy conservation, we infer $T_\text{reh}$ numerically from
\begin{align}
\frac{\pi^2}{30} g_\text{reh} T_\text{reh}^4 = \frac{\pi^2}{30} g_\text{p} T_\perc^4 + \Delta V \, .
\end{align}
A rough estimate of $T_\text{reh}$ is given by $T_\text{reh} \simeq \ba{1 + \alpha}^{1/4} T_\perc$. For the case of strong 
supercooling with $\alpha \gg 1$ one therefore finds a large hierarchy between $T_\text{reh}$ and $T_\perc$.

We note that BBN constraints imply that the reheating temperature must be larger than a
few MeV such that the SM neutrinos return to equilibrium with the thermal plasma and the
standard cosmological evolution is recovered~\cite{Hasegawa:2019jsa,Barbieri:2025moq}. We
do not implement these constraints in detail but simply adopt the conservative bound
$T_\text{reh} > 3 \, \mathrm{MeV}$ as a hard requirement for the allowed parameter space.
The preferred reheating temperatures found in our scans are sufficiently far above this
bound that its precise value does not matter for our analysis.

\paragraph{Maximal bubble size} On top of the GW signal produced right after the phase transition, 
a secondary contribution was recently identified as relevant in the limit of large,
almost Hubble-sized vacuum bubbles~\cite{Lewicki:2024ghw, Lewicki:2024sfw}. This GW contribution is sourced by 
curvature perturbations 
and appears at second order in cosmological perturbation 
theory. It can lead to the production of a second, sharp peak in the GW spectrum at frequencies $f < f_\text{p}^0$
for $\beta/H \lesssim 10$. In the same limit of large vacuum bubbles, the formation of primordial black holes through
the so-called late-bloomer mechanism~\cite{Gouttenoire:2023naa} becomes important.
We checked explicitly 
that the largest possible bubble size in our setup 
never crosses the necessary threshold for these effects to become relevant.
Likewise, our scenario evades
potential additional constraints from the CMB, 
ground-based GW observatories and lensing on such primordial black holes~\cite{Gouttenoire:2023naa, Kanemura:2024pae, Lewicki:2024sfw}.

\subsection{Comparing the predicted GW signal with PTA data}
\label{sec:PTAlike}

In order to fit the produced signal to the GW spectrum observed by PTAs, we use the 15\,yr data set released 
by the NANOGrav collaboration~\cite{NANOGrav:2023gor}. 
Given that all PTA collaborations found evidence for a GW signal with 
a compatible spectral amplitude and slope, we expect no significant impact on our results from including additional data 
sets (see section~\ref{sec:outlook}). We employ
\textsf{PTArcade}~\cite{Mitridate:2023oar} using the \textsf{ceffyl}~\cite{Lamb:2023jls} backend to compute the PTA 
likelihood $\mathcal{L}_\text{PTA}$ for a given GW spectrum. Note that the statistical biases described in 
ref.~\cite{Bringmann:2023opz}, which required the use of the computationally more intensive backend code 
\textsf{enterprise}~\cite{enterprise, enterprise2} are circumvented in the present analysis: The GW amplitudes that we 
consider always lie within the prior bounds of the individual Fourier amplitudes of the Bayesian spectrograms (``violins'') due 
to the SMBH binary contribution. 

\begin{figure}[t]
  \centering
  \includegraphics[width=.9\textwidth]{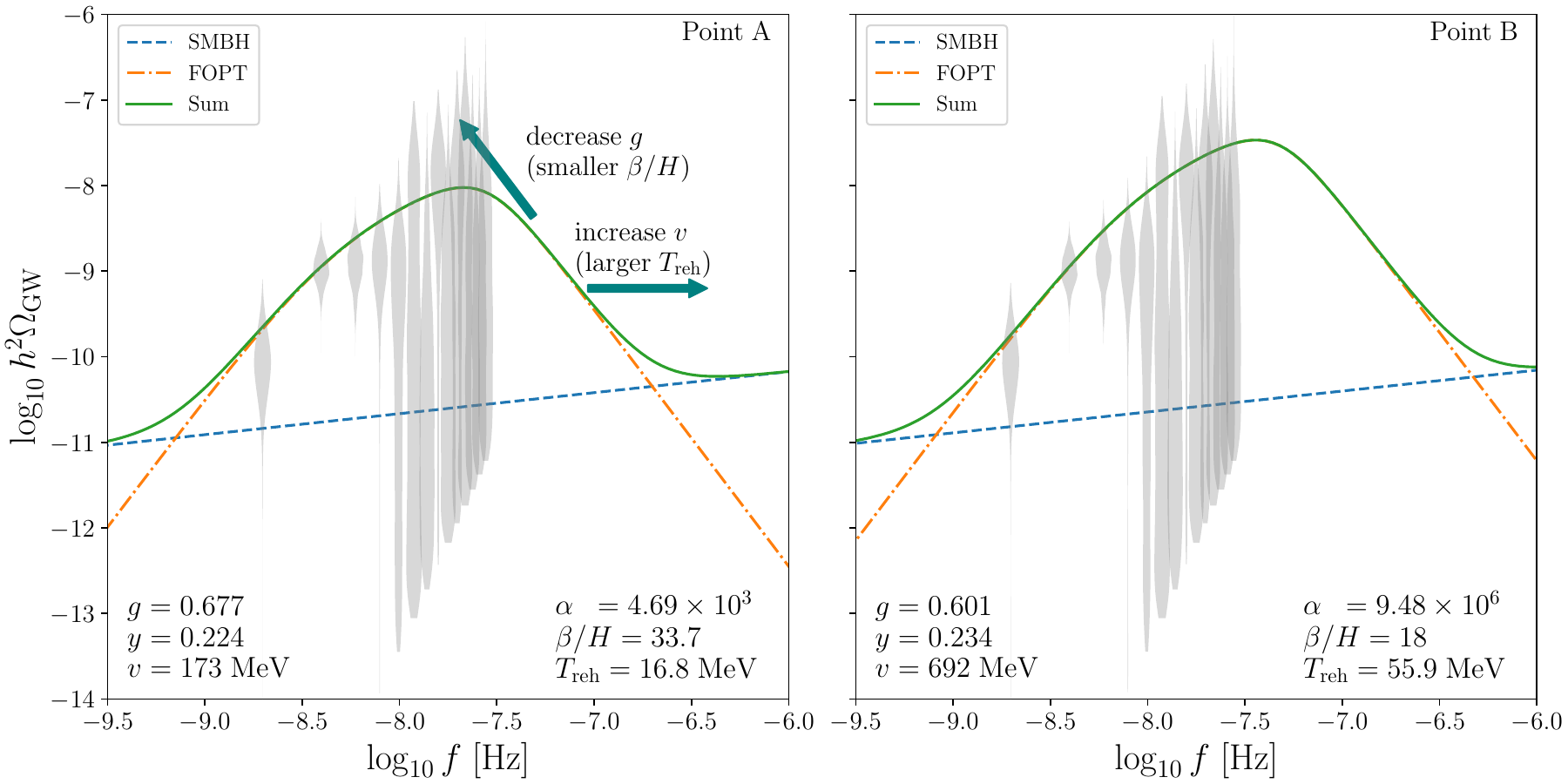}
  \caption{Gravitational wave spectra and NANOGrav 15~yr spectrograms for the two
    parameter points given in table~\ref{tab:bestfit}. The best-fit SMBH contribution to
    the GW signal for these parameter points is given by power laws with amplitude and
    slope $(\log_{10}A, \gamma) = (-15.67, 4.755)$ and $(-15.66, 4.756)$, respectively.}
  \label{fig:pta-data-signal}
\end{figure}

Figure~\ref{fig:pta-data-signal} shows the GW spectrum of the parameter points A and B
given in table~\ref{tab:bestfit}. The left panel shows the best-fit GW spectrum of the
coupled dark sector scenario. In this scenario a good fit to the PTA likelihood
$\mathcal{L}_{\mathrm{PTA}}$ can be obtained by a strong ($\alpha = 4.69\times10^{3}$) and slow
($\beta/H = 33.7$) phase transition at $T_\perc = 2.28\,\text{MeV}$. The right panel
shows the best-fit GW signal in the secluded dark sector case after additional selection
requirements. As will be explained in detail in section~\ref{sec:results}, this case
requires a larger vev, which in turn shifts the GW signal to a larger peak frequency. In
order to still obtain a good fit, the phase transition needs to be even stronger
($\alpha = 9.48\times10^{6}$) and slower ($\beta/H = 18.0$), which is achieved by a slightly lower gauge
coupling $g$ compared to the coupled dark sector scenario. Notably, the combination of
FOPT and SMBH signal give a much better fit to the GW spectrum than a single power-law
contribution from SMBH binaries in both the coupled
($\Delta \log \mathcal{L}_\text{PTA} = 6.8$) and secluded
($\Delta \log \mathcal{L}_\text{PTA} = 6.6$) dark sector scenarios.

\begin{figure}[t]
  \centering
  \includegraphics[width=0.6\textwidth]{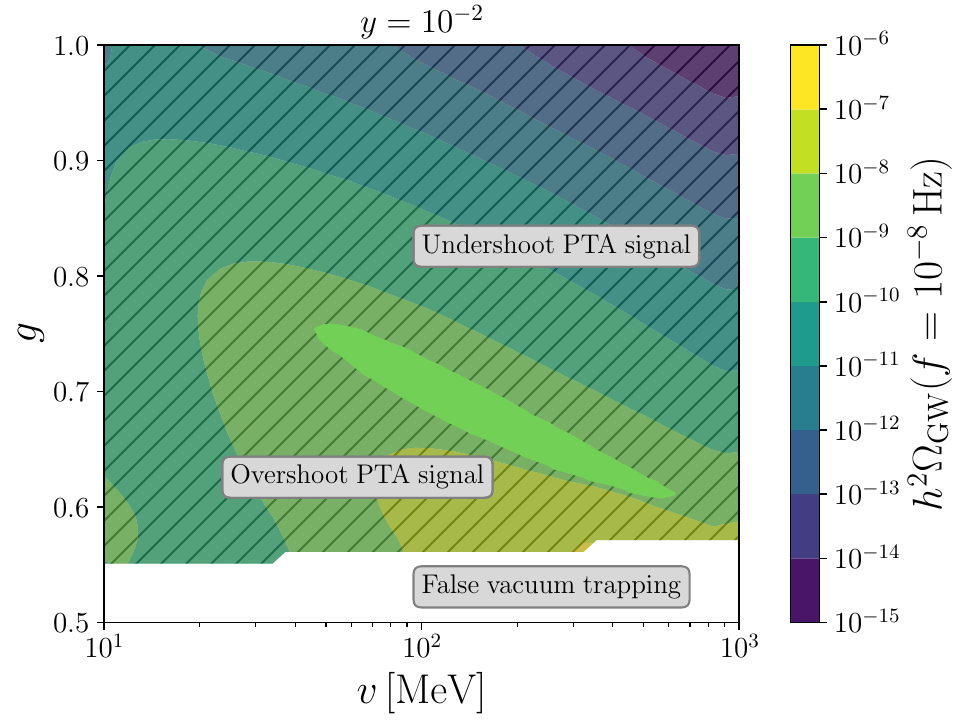}
  \caption{Amplitude and profile likelihood of the GW signal for $y = 10^{-2}$. The colour
    coding shows the amplitude of the GW signal for a fixed frequency of $f = 10^{-8}\,$Hz
    corresponding to the 5$^{\mathrm{th}}$ PTA data bin. The unshaded region corresponds to
    the 2$\sigma$ region of the PTA likelihood $\mathcal{L}_{\mathrm{PTA}}$ obtained with \textsf{PTArcade}.} 
  \label{fig:PTA-GW-likelihood}
\end{figure}

Figure~\ref{fig:PTA-GW-likelihood} shows the dependence of the FOPT-induced GW signal on the
model parameters $g$ and $v$ for a Yukawa coupling $y = 0.01$ and a fixed frequency of
$f = 10^{-8}\,\text{Hz}$, which is close to the middle of the log-spaced Fourier frequency
spectrum of the PTA. As shown in figures~\ref{fig:masses} and~\ref{fig:pt-alpha-betaH},
the dark Higgs bosons mass and the FOPT contribution to the GW signal becomes virtually
independent of the Yukawa coupling $y$ for $y \lesssim 0.1$ due to the suppressed effect of
fermion loops. The grey-shaded hatched regions correspond to a 2$\sigma$-excluded region of parameter
space.
Note  that in this plot we do not include a SMBH binary contribution.

As is visible in this figure, fitting the PTA data requires a specific combination of $g$ and $v$: The rising slope of
the GW spectrum requires the peak frequency to be larger than
$f_{\mathrm{peak}} \gtrsim 10^{-8} \, \text{Hz}$, which places a lower bound on the vev of
$v \gtrsim 50\,\text{MeV}$. Larger vevs shift the signal to higher frequencies, which has to be
compensated by a larger amplitude in order to maintain a good fit. Since smaller values of
$g$ correspond to slower transitions (lower $\beta/H$) and stronger supercooling (larger
$\alpha$), and thus larger GW amplitudes, the preferred region of the PTA likelihood decreases
in $g$ with increasing $v$. Below this preferred region the combination of a too-strong
signal and a low peak frequency exceed the PTA data. Above,  
the amplitude of the FOPT signal becomes too small, 
and an additional sizable contribution from SMBH binaries is needed to explain the data.
For $v \gtrsim 700 \, \text{MeV}$, the peak frequency becomes too large to still be compensated by an increase in the peak 
amplitude, which cannot become much larger than $h^2 \Omega_\text{GW}^\text{peak} \simeq 10^{-7}$ in our model
(due to a lower bound on the transition speed of $\beta/H \gtrsim 10$).
We note that these maximal peak amplitudes are significantly smaller than the currently strongest 
$\Delta N_\text{eff}$ constraints on the energy density carried by the GW background, cf.~eq.~(4.26) in 
ref.~\cite{Yeh:2022heq}, such that we ignore this contribution in the present work.

\section{Relic density calculation}
\label{sec:relicdensity}

Generally speaking, there are two types of processes relevant for the calculation of the
DM relic density: processes connecting the dark sector with the SM, and processes within
the dark sector. The former include DM annihilation and creation
($\chi \bar{\chi} \leftrightarrow e^+ e^-$), dark Higgs decays and inverse decays
($\phi \leftrightarrow e^+ e^-$) and DM scattering off electrons
($\chi e^\pm \leftrightarrow \chi e^\pm$), all of which rely on off-shell dark photons and depend on the kinetic
mixing parameter $\kappa$. The latter type includes in particular the conversion of DM particles
and dark Higgs bosons $\chi \bar{\chi} \leftrightarrow \phi \phi$, as well as the scattering process
$\chi \phi \leftrightarrow \chi \phi$. In principle, there could also be processes involving on-shell dark photons,
but we will see that in the interesting regions of parameter space
$m_{A^\prime} \gg m_\chi, m_\phi$, such that the dark photon contribution is strongly suppressed.

For the range of kinetic mixing parameters that we consider, the scattering of DM
particles off electrons is always efficient, i.e.~much larger than the Hubble
  rate, as long as electrons are relativistic and abundant in the plasma. 
We therefore
assume that the SM and dark sector particles share a common temperature
$T$.\footnote{%
\label{foot:phi_not_thermal}
This includes also the assumption that $\phi$ follows the same thermal distribution as $\chi$,
and hence that scattering of DM particles with dark Higgs bosons is very efficient in the temperature
ranges of interest. In the `forbidden' annihilation regime with $m_\phi\gtrsim m_\chi$~\cite{DAgnolo:2015ujb}
this is however typically not fully satisfied, which may lead to an underestimation of the DM relic density by a 
factor of a few when solving the standard Boltzmann equation~\cite{Binder:2021bmg,Aboubrahim:2023yag}.
Two comments are in order, justifying our approach even in this regime: 
{\it i)} we expect a much smaller effect than in the scenarios
studied in ref.~\cite{Binder:2021bmg,Aboubrahim:2023yag}, because in our case it is the DM particle $\chi$ itself 
(not $\phi$) that is efficiently kept in kinetic equilibrium with the heat bath; {\it ii)} solving the (numerically quite
challenging) full Boltzmann equations at phase-space level would
only lead to logarithmic corrections of the couplings needed to obtain the correct relic density, since $\Omega h^2$ depends exponentially on the DM annihilation rate in the forbidden
regime; this
would hardly be visible in the best-fit regions constituting our main results. 
}  
    
The relic abundance of DM particles is then set by the processes that change the DM
number density, i.e.~$\chi \bar{\chi} \leftrightarrow e^+ e^-$ and
$\chi \bar{\chi} \leftrightarrow \phi \phi$. We provide the full cross sections in appendix~\ref{app:freeze-out}.
The freeze-out of DM annihilations into SM fermions can be calculated using standard
methods, since electrons and positrons are in thermal equilibrium with photons and have
negligible chemical potential in the relevant temperature range. 
 For the annihilation of DM particles into dark Higgs bosons,
however, this only holds if the latter remain in equilibrium with the SM heat bath
  during the freeze-out process.
Since the loop-induced decays are not sufficient  to keep the dark Higgs bosons in chemical 
equilibrium with the SM thermal bath, cf.~the discussion around eq.~(\ref{eq:tauphi}),
this leaves us with the two possibilities introduced
as \emph{coupled dark sector} and $\emph{secluded dark sector}$, respectively,
in section \ref{sec:SM_interactions}.
For a given parameter point, the first option will always result in a smaller DM relic density
than the second. For all relic density calculations we use \ds~\cite{Bringmann:2018lay}, via the
 interface to {\sf GAMBIT} provided by {\sf DarkBit}~\cite{GAMBITDarkMatterWorkgroup:2017fax}.

\subsection{Coupled dark sector}

If dark Higgs boson decays are fast compared to the expansion of the universe, we can
assume vanishing chemical potentials for the dark Higgs bosons. In this case, it is
straight-forward to calculate the freeze-out of the annihilation process
$\chi \bar{\chi} \to \phi \phi$. As long as $m_\chi > m_\phi$, the annihilations are very efficient 
because this hierarchy in our model implies large Yukawa couplings $y\gtrsim0.3$.
The DM density will thus be strongly depleted before the DM particles freeze out, such that the final
relic abundance will be orders of magnitude below the observed value.
 In order to match observations, it therefore becomes necessary to have
$m_\chi < m_\phi$, such that annihilations are forbidden at zero temperature and
Boltzmann-suppressed during freeze-out~\cite{DAgnolo:2015ujb}.

\subsection{Secluded dark sector}
\label{sec:dec_dark_sector}

\begin{figure}[t]
  \centering
  \includegraphics[width=0.6\textwidth]{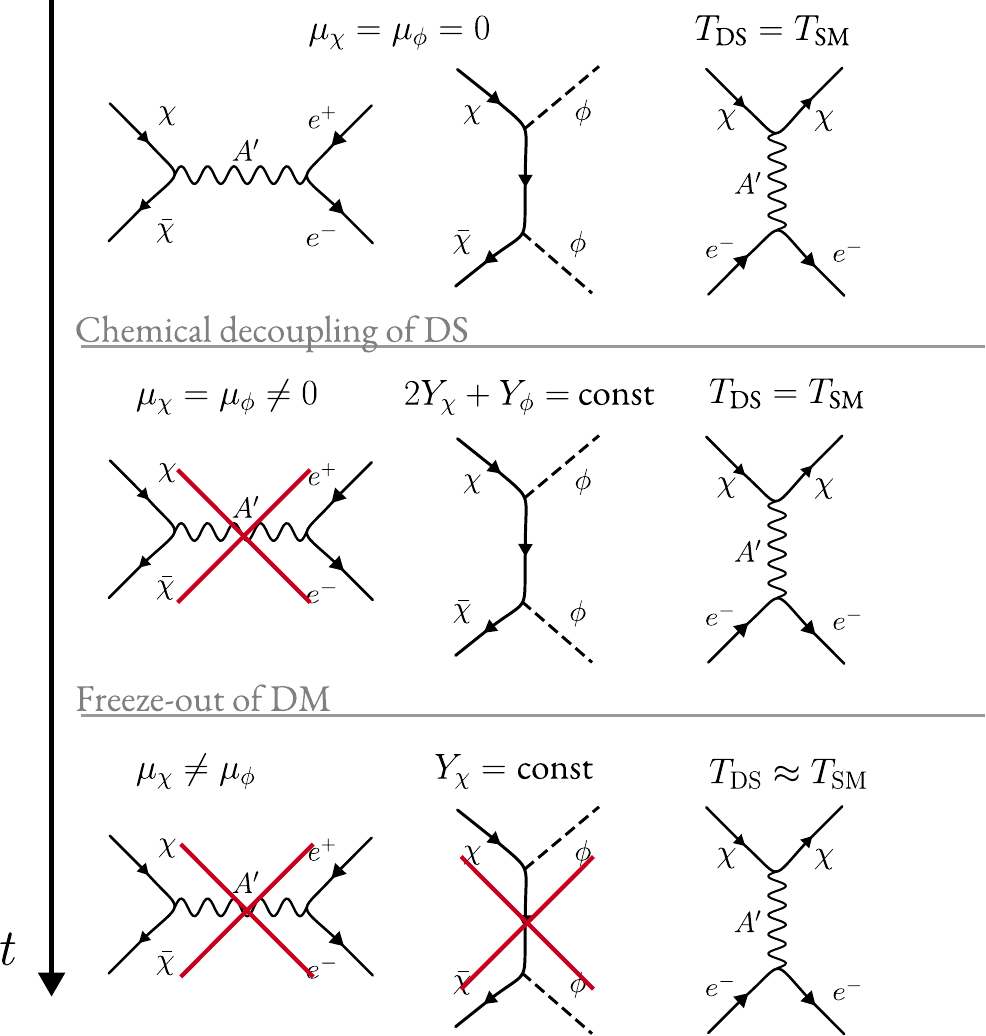}
  \caption{Relevant processes for the thermalisation and subsequent decoupling of SM electrons, the DM particle $\chi$ and the dark Higgs boson $\phi$ in the {\it secluded dark sector} scenario.}
  \label{fig:thermalisation-diagrams}
\end{figure}

The evolution of a secluded dark sector
is illustrated in figure~\ref{fig:history}, which we complement in figure~\ref{fig:thermalisation-diagrams}
with a more detailed description in terms of the underlying interaction processes.
At high temperatures, when the
dark sector is still in equilibrium with the SM, the chemical potential of all
particle species vanishes (top panel). Once the process $\chi \bar{\chi} \to e^+ e^-$ has dropped out
of chemical equilibrium, there are no efficient number-changing processes in the dark
sector (middle panel),\footnote{%
  This necessarily holds if the dark sector decouples when already non-relativistic. 
  Otherwise there might still be efficient number-changing processes, such as $3\phi\to2\phi$ and
  $\chi \bar{\chi} \to 3 \phi$~\cite{Farina:2016llk,Cline:2017tka}, a detailed study of which is left
  for future work. 
  We note, however, that in such a case the dark sector would initially still behave {\it as if}  in chemical
  equilibrium with the SM, because the dark sector interactions would enforce a vanishing
  chemical potential (while $\chi e^\pm \leftrightarrow \chi e^\pm$ keeps kinetic equilibrium 
  with the SM). However, the number-changing processes are expected to
  become inefficient before the process $\chi \bar{\chi} \to \phi \phi$ freezes out, 
  see also ref.~\cite{Bringmann:2020mgx}, such that chemical
  potentials will eventually become non-negligible.
    } 
such that the sum of DM particles,
DM antiparticles and dark Higgs particles remains constant:
$N_\chi + N_{\bar{\chi}} + N_\phi = \text{const}$. Introducing the number densities
$n_{\chi,\phi}$, the SM entropy density $s(T)$ and the yields
$Y_{\chi,\phi} = n_{\chi,\phi} / s$, this equation can also be written as
\begin{equation}
Y_\chi + \tfrac{1}{2} Y_\phi = \text{const} \,.
\label{eq:constant_number}
\end{equation}
In practice, we determine this constant at the temperature $T_f^\text{SM}$ when the dark sector decouples from the
SM, i.e.~at the freeze-out temperature of the process $\chi\chi\leftrightarrow e^+e^-$.

For $T<T_f^\text{SM}$, as long as the DM particles and the dark Higgs bosons are in equilibrium with each other, 
detailed balance requires
\begin{equation}
    Y_\chi^2 \langle \sigma v \rangle_{\chi\chi\to\phi\phi} = Y_\phi^2 \langle \sigma v \rangle_{\phi \phi \to \chi \chi} \, ,
\label{eq:detailed_balance}
\end{equation}
where the brackets denote thermal averages.
Together, eqs.~\eqref{eq:constant_number} and~\eqref{eq:detailed_balance} determine the number densities of DM 
particles and dark Higgs bosons as a function of temperature, see appendix~\ref{app:freeze-out} for further details.
This, along with  the thermally averaged cross sections, allows us to 
determine the temperature $T_f^\phi$ at which the dark sector 
interactions eventually freeze out (bottom panel of figure~\ref{fig:thermalisation-diagrams}). 
At this point, annihilation and production processes become negligible, such that for 
$T < T_f^\phi$ we find $Y_\chi(T) = Y_\chi(T_f^\phi)$ and $Y_\phi(T) = Y_\phi(T_f^\phi)$. For the case of DM particles, 
this yield directly translates to the present-day yield $Y_{\chi,0}$. The dark Higgs boson abundance, on the other
hand, must eventually decay -- leading to additional constraints discussed in the next section. 

While we refer to the appendix for a more detailed discussion, let us note that we can distinguish two qualitatively
different regimes for the evolution of a secluded dark sector.
For $\Delta m\equiv m_\phi - m_\chi\gtrsim0$, 
the energy density of the dark sector is largely dominated by DM particles at 
$T = T_f^\text{SM}$, by a factor of roughly $\exp(\Delta m/ T_f^\text{SM})$, and any interactions between DM 
particles and dark Higgs bosons will only further deplete the energy density of the latter 
(because kinematics strongly 
favour $\phi\phi\to\chi\chi$ over $\chi\chi\to\phi\phi$). Hence, the DM abundance is, to a good approximation,
determined primarily by the freeze-out of annihilations into SM particles, $Y_{\chi,0} \approx Y_\chi(T_f^\text{SM})$. If 
the DM particles are heavier ($\Delta m\lesssim0$), on the other hand, the annihilation of DM particles into dark Higgs 
bosons in the temperature range $T_f^\text{SM}>T>T_f^\phi$ will significantly deplete the initial abundance of DM 
particles. 

For both the coupled and the secluded dark sector, we compare the calculated DM relic
density to the value $\Omega h^2 = 0.120 \pm 0.001$ measured by Planck~\cite{Planck:2018vyg}. We
note that, in our relic density calculation, we have a sizeable theory uncertainty that is
much larger than the measurement uncertainty: for the coupled dark sector case, this is
due to the effects of early kinetic decoupling not included in our calculation. For the
secluded dark sector, the main uncertainty comes from the sudden freeze-out approximation
for the $\phi\phi\to\chi\chi$ annihilations. We therefore include a factor of 2 uncertainty in the relic
density prediction, which is implemented as a Gaussian likelihood:
\begin{equation}
\log_2 \left(\frac{\Omega h^2}{0.12}\right) = 0 \pm 1 \; .
\end{equation}
Let us stress, however, that the relic density depends
exponentially on the model parameters in the critical regimes; this implies that even a
factor 2 uncertainty in the relic density calculation translates into quite modest shifts
in the underlying parameters.

\section{Constraints}
\label{sec:constraints}

The most prominent phenomenological features of the model that we study are the production of GWs from a FOPT, which 
are constrained by the PTA and SMBH binaries likelihoods as discussed in section~\ref{sec:gw}, and the freeze-out of DM, 
which is constrained by the observed relic density as discussed in section~\ref{sec:relicdensity}. The same processes 
that are relevant for these features can occur at other times in the cosmological history or under laboratory conditions,
leading
to further constraints relevant for our model. 

\subsection{Cosmological constraints}
\label{sec:cosmo_constraints}

For our study, we focus on  
dark sector particles in the mass range 
$\mathcal{O}(10-100)\,\mathrm{MeV}$. If these particles remain in chemical equilibrium with the SM thermal bath, their 
number densities are strongly Boltzmann-suppressed during BBN 
and do not 
directly affect primordial element abundances or the CMB. The only relevant effect in this case is a slight modification of the 
Hubble rate during neutrino decoupling, which can change the helium abundance, effectively imposing the bound 
$m_\chi \gtrsim 10 \, \mathrm{MeV}$~\cite{Sabti:2019mhn}. This effect is implemented in {\sf AlterBBN}~\cite{Arbey:2018zfh}, 
which we use for the 
calculation of the primordial element abundances, see ref.~\cite{GAMBITCosmologyWorkgroup:2020htv} for a description 
of the {\sf CosmoBit} interface as well as ref.~\cite{Balazs:2022tjl} for the implemented likelihoods of the primordial element 
abundances.

If, on the other hand, the dark sector decouples from the SM thermal bath before or during BBN, out-of-equilibrium 
annihilations or decays can change the cosmological history in more complex ways~\cite{Hufnagel:2018bjp, Kawasaki:2020qxm}. In particular 
the decays of decoupled dark Higgs bosons are constrained by various observables that depend on the abundance, mass 
and lifetime of dark Higgs bosons. If the dark Higgs bosons decay into electrons after neutrino decoupling, the injected 
particles thermalise with the SM bath and increase the photon temperature, which decreases  both the baryon-to-photon ratio 
$\eta$ (due to the entropy injection) and the number of relativistic degrees of freedom 
$N_{\mathrm{eff}}$ (by effectively reducing the relative neutrino temperature).

In our analysis, we use {\sf AlterBBN} to calculate the change of $N_{\mathrm{eff}}$ due
to DM annihilations and dark Higgs boson decays during BBN, taking an initial value of
3.044~\cite{Bennett:2019ewm,Akita:2020szl,Bennett:2020zkv,Froustey:2020mcq}. The resulting value of $N_\text{eff}$ is then compared to the Planck
likelihood \cite{Planck:2018vyg}.
\begin{equation}
  N_{\mathrm{eff}} = 2.99 \pm 0.17
\end{equation}
which combines TT,TE,EE+lowE+lensing+BAO.

For the baryon-to-photon ratio $\eta$, we proceed in the opposite direction. We assume that
this value is fixed at late times through CMB measurements,
as $\eta_\text{CMB} = 6.117 \cdot 10^{-10}$,
and then numerically solve the Boltzmann equation for
the dark Higgs boson energy density as well as the Friedmann equation, allowing for an
intermediate period of matter domination followed by sizable entropy injection, see
ref.~\cite{Balazs:2022tjl} for details. As result of this calculation we obtain the
initial value $\eta_\text{BBN}$ needed to reach the CMB value after entropy injection. This
value is then used as an input parameter for {\sf AlterBBN} and affects the primordial element
abundances. The shift between $\eta_\text{CMB}$ and $\eta_\text{BBN}$ is found to give the
strongest bound on the dark Higgs abundance for dark Higgs lifetimes of the order of
$10^{3}\,\mathrm{s}$, while for shorter lifetimes the change of $N_\text{eff}$ has a
larger effect~\cite{Hufnagel:2018bjp}.

For $\tau_{\phi}>10^4 \, \mathrm{s}$, on the other hand, strong constraints come from the photodisintegration of light 
nuclei~\cite{Depta:2020zbh}. 
At these times, the energy injected by dark Higgs boson decays is no longer efficiently converted into low-energy photons 
via electromagnetic cascades.
Instead, it may dissociate the light nuclei 
formed during BBN, in particular deuterium. We study the effect of photodisintegration using the public code 
{\sf ACROPOLIS}~\cite{Depta:2020mhj,Balazs:2022tjl}, which takes as input the predicted abundances from {\sf AlterBBN} and 
calculates the further evolution. 

For even longer lifetimes, dark Higgs boson decays are strongly constrained by measurements of CMB spectral distortions 
($\tau_\phi \gtrsim 10^{7} \, \mathrm{s}$) and  the CMB power spectra ($\tau_\phi \gtrsim 10^{12}\,\mathrm{s}
$)~\cite{Poulin:2016nat}. We find that the allowed parameter regions in our scans correspond to lifetimes 
$\tau\ll10^{7}\,\mathrm{s}$, and hence do not expect these constraints to be relevant
for us.

\subsection{Direct and indirect detection constraints}

In the model that we consider, both annihilations of DM particles into SM fermions and annihilations into dark Higgs bosons 
proceed via $p$-wave and are therefore strongly suppressed in the non-relativistic limit~\cite{Battaglieri:2017aum}. As a 
result, there are no relevant constraints on DM annihilations from either CMB observations or astrophysical systems, which 
are highly relevant for sub-GeV DM models with $s$-wave annihilation~\cite{Balan:2024cmq}. 
For the same reason, late-time DM annihilations cannot affect the light element abundances.
We note that the process 
$\chi \bar{\chi} \to \phi A'$ would in principle be observable~\cite{Bell:2016fqf}, but is kinematically forbidden in the 
parameter regions of interest.

Dark photon exchange gives rise to the scattering of DM particles off SM fermions via the effective interaction
\begin{equation}
\mathcal{L}_\text{eff} = \frac{1}{\Lambda^2} \bar{\chi} \gamma^\mu \gamma^5 \chi \bar{f} \gamma_\mu f \, .    
\end{equation}
The resulting scattering cross section is momentum-suppressed in the non-relativistic limit. In combination with the general 
difficulties of detecting the scattering of sub-GeV DM particles, this suppression renders constraints from direct detection 
experiments irrelevant for the model that we consider. 
We therefore do not include any direct detection likelihoods in our analysis.

\subsection{Dark photon searches}

In the entirety of the parameter space considered, we have $g \gg \kappa$. The dark photon thus predominantly decays 
into DM particles. Hence, we include constraints on invisibly decaying dark photons from 
LSND~\cite{LSND:2001akn,deNiverville:2011it}, MiniBooNE~\cite{PhysRevD.98.112004}, NA64~\cite{NA64:2023wbi} and 
BaBar~\cite{BaBar:2017tiz} as implemented in ref.~\cite{Balan:2024cmq}. LSND and MiniBooNE are beam dump 
experiments in which neutral mesons are produced from proton beams dumped onto a dense target. The neutral mesons 
can then decay into photons and dark photons, which eventually decay into DM particles. A downstream detector looks for 
these DM particles through DM-electron or DM-nucleon scattering. The number of signal events, $s \propto \kappa^4 g^2$, 
is calculated using BdNMC~\cite{deNiverville:2016rqh}. The signal prediction is passed onto a Poisson likelihood with $n$ 
observed and $b$ SM background events.

In NA64, an electron beam is dumped onto a fixed target, possibly producing dark photons via dark bremsstrahlung.  We 
then look for events with missing energies taken away by the dark photons (which for the gauge couplings that we
consider decay essentially immediately to invisible DM particles). The predicted 
number of missing energy events, $s \propto \kappa^2$ is used to calculate a Poisson likelihood as discussed before. 

BaBar looks for invisibly decaying dark photons produced along with mono-energetic photons
in $e^+e^-$ collisions. We use a likelihood function that reproduces limits from
ref.~\cite{BaBar:2017tiz} on the kinetic mixing parameter as a function of the dark photon
mass. This likelihood excludes $\kappa > 10^{-3}$ in the entire dark photon mass range
under consideration, while NA64 gives even stronger bounds for
$m_{A'} \lesssim 300 \, \mathrm{MeV}$. For $g < 1$, the constraints from LSND and MiniBooNE are
found to be subleading. This is also expected to hold true for the recent MicroBooNE
result~\cite{MicroBooNECollaboration:2023kmx}, which is difficult to reinterpret and
hence not included in our analysis.

\subsection{Dark matter self-interactions}
The DM particles in this model undergo self-interactions in the form of both particle-particle and particle-antiparticle 
scattering mediated by dark photon and dark Higgs boson exchange. Self-interactions can be studied in astrophysical 
systems on a wide range of scales from satellite galaxies to galaxy clusters. The most stringent constraints come from 
galaxy clusters that have characteristic velocities of order $10^3\,\mathrm{km/s}$. Thus, standard astrophysical systems 
only probe the scattering cross sections in the non-relativistic limit.\footnote{%
We note in passing that DM around supermassive black holes may form dense spikes where relativistic velocities can be 
reached \cite{Gondolo:1999ef,Shapiro:2014oha}, providing an opportunity to probe DM self-interactions in a different 
regime.
} 
In our model, self-interactions are dominated by a velocity-independent term in the non-relativistic limit. 
We also safely remain within the Born regime, $(g^2/4\pi) m_\chi\ll m_\phi$,  where cross sections
can be calculated perturbatively~\cite{Feng:2009hw}.
These cross sections 
are given by
\begin{align}
\sigma_{\chi\chi \to \chi\chi} & = \frac{y^2}{256 \pi v^2} \left(3 - 4 \frac{m_\chi^2}{m_\phi^2}\right)^2 \\
\sigma_{\chi\bar{\chi} \to \chi\bar{\chi}} &= \frac{y^2}{128 \pi v^2}\left(7 - 4 \frac{m_\chi^2}{m_\phi^2} + 
 16 \frac{m_\chi^4}{m_\phi^4}\right),
\end{align}
and constraints on an isotropic, velocity-independent cross-section can be directly applied. 

The Bullet Cluster consists of a small subcluster passing through a larger main 
cluster~\cite{Clowe:2003tk,Clowe:2006eq,Bradac:2006er,Paraficz:2012tv}. Scatterings of DM particles from the subcluster 
of those from the main cluster can lead to ejection of particles from the subcluster. The fraction of mass lost by the 
subcluster can then be used to constrain the self-interaction cross-
section~\cite{Markevitch:2003at,Randall:2008ppe,Kahlhoefer:2013dca}. The mass loss is determined from the observed 
subcluster mass-to-light ratio (MLR) and an assumed initial MLR. We treat the initial MLR as a nuisance parameter and 
calculate a marginalised likelihood for the DM parameters. For the initial subcluster MLR, we use a Gaussian prior centered 
around the measured main cluster MLR, see ref.~\cite{Balan:2024cmq} for details.

\section{Results}
\label{sec:results}

As discussed in section~\ref{sec:relicdensity}, our analysis considers two
phenomenologically distinct cases. First, we consider a dark sector that remains in
thermal equilibrium with the bath of SM particles throughout the phase transition and the
freeze-out of DM. We refer to this scenario as the \textit{coupled dark sector} setup, as
opposed to the \textit{secluded dark sector} scenario, in which the dark sector decouples
from chemical equilibrium with the SM bath after the phase transition but before DM
freeze-out. The main difference between the two cases is that the latter features a
population of decoupled dark Higgs bosons that decay during or after BBN, leading to
additional constraints on the model (see section~\ref{sec:constraints}). As we will see,
the two scenarios therefore favor different regions of model parameter space.

Before presenting our main results, let us briefly review the model parameters and their
ranges. The dark sector coupling $g$ can in principle vary by orders of magnitude up to
the perturbativity bound $g < \sqrt{4\pi}$. In practice, however, only a narrow range of
$g$ yields the strong FOPT needed to fit the PTA signal, see the discussion in
section~\ref{sec:PTAlike}. In our scans we therefore restrict ourselves to
$0.3 \leq g \leq 1$. For the same reason, we require the dark Higgs vev to lie in the range
$10 \, \mathrm{MeV} < v < 1 \, \mathrm{GeV}$. As discussed in section~\ref{sec:gw},
we do not consider values of the vev larger than $1\,\mathrm{GeV}$, for which the FOPT
would have to be very slow ($\beta / H < 10$) in order to fit the PTA signal. The upper bound
on $v$ also ensures that $m_\phi < 200 \, \mathrm{MeV}$, so that we do not have to consider
decays of the dark Higgs boson into muons and pions.

The DM mass turns out to be quite tightly constrained by the relic density requirement.
For the case of a coupled dark sector, we vary $0.15 < y < 0.45$, which corresponds
roughly to $0.5 \lesssim m_\chi / m_\phi \lesssim 1$, i.e.\ the regime of forbidden annihilations. For the
secluded dark sector, on the other hand, we find $m_\chi \approx m_\phi$ (see
appendix~\ref{app:freeze-out}), such that it turns out to be convenient to introduce the
parameter
\begin{equation}
    \delta = \frac{m_\chi - m_\phi}{m_\chi} \, ,
    \label{eq:delta}
\end{equation}
which we vary in the range $10^{-3} \leq \delta \leq 0.1$.

Values of the kinetic mixing parameter larger than $10^{-3}$ are robustly excluded by
experiments, while values smaller than $10^{-7}$ are completely irrelevant for
phenomenology, such that we choose these two values as upper and lower bound for the
sampling range of $\kappa$ respectively. For the secluded dark sector, somewhat larger values
of $\kappa$ are required to sufficiently deplete the dark sector, such that we restrict our
scan ranges to $10^{-5} \leq \kappa \leq 10^{-3}$.

The exploration of the parameter space is done with the {\sf GAMBIT} global fitting
framework, using \textsf{Diver} with a population size of 38,000 and a convergence
threshold of $10^{-5}$. We have checked that these settings are sufficient to ensure that
the scans converge. The results of all scans along with relevant {\sf GAMBIT}
configuration files and example plotting scripts are available on
Zenodo~\cite{balan_2025_15863201}.

We will start by discussing the less involved \textit{coupled scenario}, before going over to the more constrained \textit{secluded dark sector} scenario.

\subsection{Coupled dark sector}
\label{sec:results_coupled}

Let us start the discussion of our results with a brief summary of the main features of the coupled dark sector scenario. 
As pointed out in  section~\ref{sec:gw}, 
for sufficiently small values of the Yukawa coupling $y$, the GW signal mainly depends on two parameters: the gauge 
coupling $g$ and the vev $v$. We already showed in figure~\ref{fig:PTA-GW-likelihood} that the PTA observations hence 
favor specific combinations of $g$ and $v$, while the DM relic density depends on the process $\chi \bar{\chi} \to \phi \phi$ 
and hence on the Yukawa coupling $y$. The kinetic mixing parameter $\kappa$ plays no significant role in the coupled dark 
sector scenario, as it does not modify the effective potential (see section~\ref{sec:conformal-models}) and thermal 
equilibrium between the dark and visible sectors is established through the assumed effective interactions between dark 
Higgs bosons and electrons (see appendix~\ref{app:toy_model}).

\begin{figure}[t]
    \centering
    \includegraphics[width=0.48\linewidth]{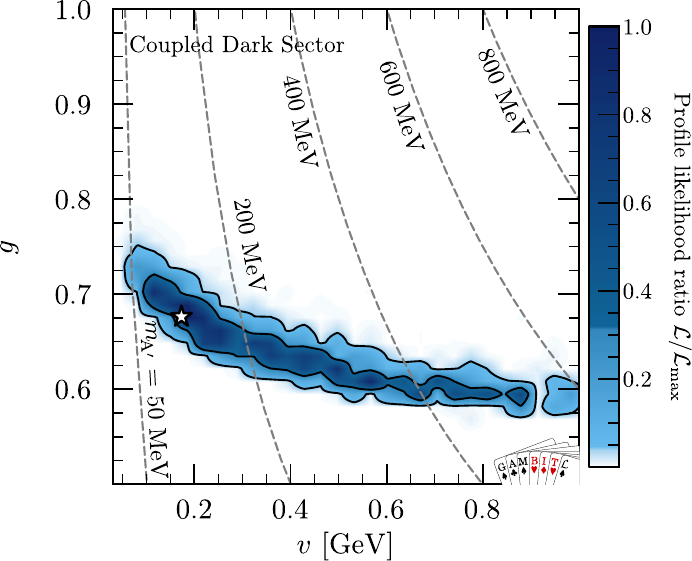}
    \hfill
    \includegraphics[width=0.48\linewidth]{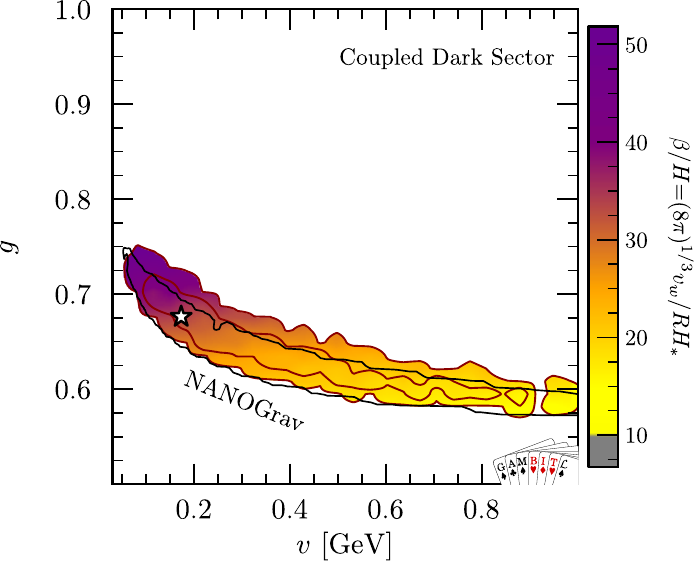}
    
    \includegraphics[width=0.48\linewidth]{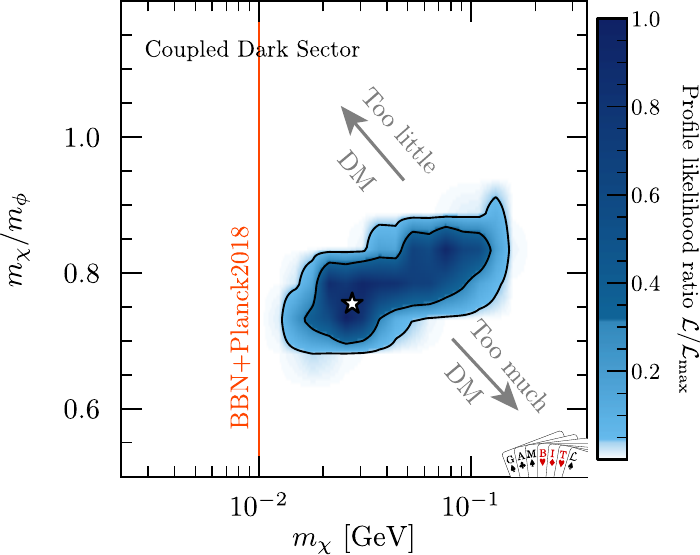}
    \hfill
    \includegraphics[width=0.48\linewidth]{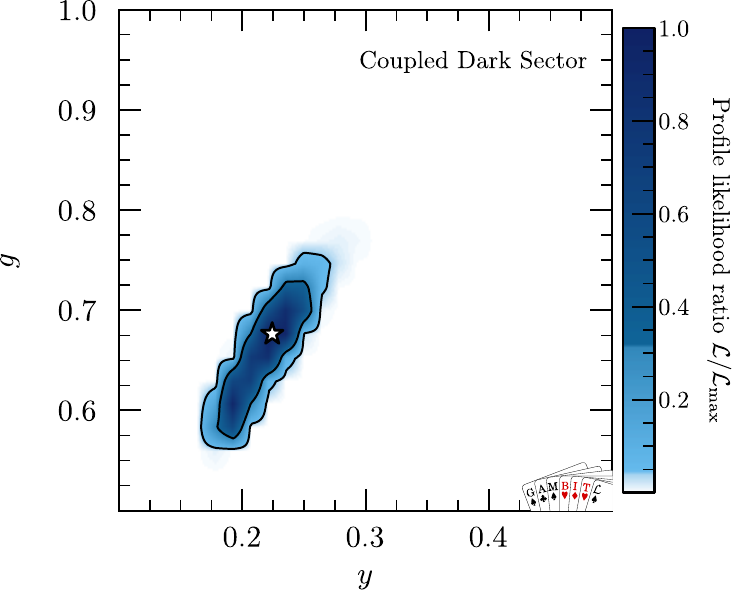}
    
    \caption{Allowed parameter regions determined through a global analysis of the coupled dark sector scenario in terms of 
    the profile likelihood (blue shading). In the top-left panel, dashed lines indicate constant values of the dark photon mass 
    $m_{A'} = g v$. In the top-right panel, we instead shade the allowed parameter region according to the speed of the 
    phase transition, parametrised by $\beta/H$. The black solid line indicates the parameter region preferred by PTA data alone when neglecting any potential SMBH contribution, 
    see figure~\ref{fig:PTA-GW-likelihood}. In the bottom-left panel we indicate the additional constraints that define the 
    allowed regions of parameter space.
}
    \label{fig:coupledDS}
\end{figure}

The allowed parameter regions for the coupled dark sector are shown in
figure~\ref{fig:coupledDS} in terms of the profile likelihood ratio with respect to the
best-fit point (called point A in table~\ref{tab:bestfit}). In the top row we consider the
\plane{g}{v} plane, which is directly constrained by the PTA signal. The gauge coupling
$g$ determines the amount of supercooling before the breaking of the $U(1)'$ symmetry,
while the vev $v$ sets the overall temperature scale and hence the frequency range. As
shown in figure~\ref{fig:PTA-GW-likelihood}, gauge couplings
$g\simeq 0.57 \text{--} 0.75$ can give rise to sufficiently large bubbles (or equivalently,
sufficiently slow phase transitions) to account for the peak amplitude of the PTA signal
($h^2 \Omega_\GW^\peak \simeq 10^{-8}$), while vevs
$v \simeq 60 \text{--} 900 \, \text{MeV}$ are required in order for the signal to peak in the
correct frequency range ($f_\peak \simeq 10^{-8} \, \mathrm{Hz}$).

The top-right panel of figure~\ref{fig:coupledDS} also shows the \plane{g}{v} parameter
plane, but now the shading indicates the speed of the phase transition $\beta/H$ instead of
the likelihood ratio as in the top-left panel. In addition, we use a black line to show
the parameter region favoured by fitting the FOPT signal to PTA data alone, neglecting
also any potential SMBH contribution (see figure~\ref{fig:PTA-GW-likelihood}). We find
typical values in the range $\beta/H \simeq 15\text{--}45$, corresponding to significant
supercooling and large bubble sizes. The percolation temperature at the best-fit point is
around $T_\perc \sim 2 \, \mathrm{MeV}$ and therefore about two orders of magnitude smaller
than the vev. For a given value of $v$, the lower bound on $g$ stems from the requirement
not to produce a GW signal larger than the one observed by PTAs, which is strongly
disfavoured and hence gives a sharp boundary. For large values of $g$, on the other hand,
the GW signal from a FOPT is too weak to explain the PTA signal, so an additional SMBH
binary signal is needed. As the latter is described by a broad distribution of spectra
with amplitudes varying over several orders of magnitude, the upper limit on $g$ is softer
and is more scattered beyond the shown 2$\sigma$ PTA contour.

With both $g$ and $v$ being determined through the PTA data, also the mass of the dark
photon is determined to lie in the range
$m_{A^\prime} \simeq 50 \text{--} 600 \, \text{MeV}$. The dark Higgs mass in principle depends
on all three parameters $g$, $v$ and $y$, see figure~\ref{fig:masses}, but for
$y^4 \ll g^4$ the dependence on $y$ is very weak and we find approximately
$m_\phi \approx 0.46 g^2 v = 0.46 g m_{A'}$. Hence the dark Higgs boson mass is roughly in the
range $m_\phi \simeq 20\text{--}170 \, \mathrm{MeV}$.

Let us now turn to the Yukawa coupling $y$, which determines the DM mass via
$m_\chi = y v / \sqrt{2}$ and the DM relic abundance. For $y \gtrsim 0.65 g^2$, the DM particle
would be heavier than the dark Higgs boson. Given the large Yukawa coupling required, the
annihilation process $\chi \bar{\chi} \to \phi \phi$ would then be extremely efficient and strongly
deplete the DM relic abundance. To reproduce the observed DM relic abundance, it therefore
becomes necessary to consider the case $m_\chi < m_\phi$, such that annihilations into dark
Higgs bosons are forbidden in the non-relativistic limit. In this case, the DM relic
abundance depends roughly exponentially on the ratio $(m_\phi - m_\chi) / T_\text{f}$, with the
freeze-out temperature $T_\text{f}$. Smaller DM masses therefore correspond to a larger DM
relic abundance. In practice, we find that the observed DM relic abundance implies a mass
ratio $m_\chi / m_\phi$ between 0.7 and 0.9, see the bottom-left panel of
figure~\ref{fig:coupledDS}. This, in turn, determines the Yukawa coupling to
$y \approx 0.15 \text{--} 0.3$, with a tight positive correlation with $g$, see the
bottom-right panel of figure~\ref{fig:coupledDS}. This correlation results from the
requirement $m_\chi \approx 0.75 m_\phi$, which translates to $y \approx 0.5 g^2$.

With the observationally favored region of parameter space precisely determined, let us
briefly discuss the corresponding thermal history of the dark sector, taking for
concreteness the values of the best-fit point. The spontaneous breaking of the $U(1)'$
symmetry and the breaking of conformal symmetry occur through the nucleation of vacuum
bubbles at a temperature of $2.8 \, \text{MeV}$, which subsequently percolate at
$2.3 \, \text{MeV}$. A sizeable reheating of the combined plasma of dark sector and SM
particles then increases the temperature of the plasma back to $16.8 \, \text{MeV}$. At
this temperature, the dark sector states are all non-relativistic
($m_{A^\prime} = 117 \, \text{MeV}$, $m_{\chi} = 27 \, \text{MeV}$,
$m_{\phi} = 36 \, \text{MeV}$) and thus their number density becomes exponentially suppressed
due to the rapid decays of dark Higgs bosons into electron-positron pairs and the
efficient conversion of dark sector states into each other.
DM freeze-out happens at $T_\text{f} = 1.2 \, \text{MeV}$ at parameter point A
(corresponding to $x_\text{f} = m_\chi /T_\text{f} \approx 22$), roughly at the same temperature as
the freeze-out of neutron-proton conversion processes, which marks the beginning of BBN.
However, at this point the abundance of all dark sector states is already so strongly
suppressed that it does not affect any of the subsequent cosmological evolution. At the
best-fit point, we find $ N_\text{eff} = 3.023$, only slightly different from the
standard value.

To conclude the discussion of the coupled dark sector, we point out that the model in
principle permits for a second way to set the DM relic abundance, namely via the process
$\chi \bar{\chi} \to A'^\ast \to e^+ e^-$. This process is expected to dominate for
$m_\chi < 0.7 m_\phi$ when forbidden annihilations become strongly suppressed. However,
experimental upper bounds on the kinetic mixing parameter $\kappa$ imply that the annihilation
cross section can never be large enough to reproduce the observed DM relic abundance via
freeze-out, since the dark photon mass is well above the resonance at
$m_{A'} \approx 2m_\chi$~\cite{Balan:2024cmq}. For example, for the best-fit values of $g$ and
$v$ and a DM mass of 20 MeV (instead of 27 MeV), the observed DM relic abundance
could be reproduced for $\kappa \sim 2\times10^{-3}$. The null result from NA64, however, imposes the
upper bound $\kappa \sim 2\times10^{-4}$ for $m_{A'} \approx 120 \, \mathrm{MeV}$ and therefore rules out this
possibility. While the parameters $g$, $y$ and $v$ can hence be inferred from
observations, the kinetic mixing $\kappa$ remains largely unconstrained in the coupled dark
sector scenario and can potentially be unobservably small.

\subsection{Secluded dark sector scenario}
\label{sec:results_secluded}

We now turn to the case of a secluded dark sector, in which the kinetic mixing of the dark
photon is the only portal coupling connecting the two sectors. Since the kinetic mixing is
irrelevant for the FOPT, the discussion of the GW signal and the PTA likelihood is
essentially identical to the previous case. As before, in particular, the dark sector
particle masses in the new phase are significantly larger than the relevant temperature
scales, such that all dark sector particles are non-relativistic after reheating.

Nevertheless, immediately after the phase transition, the Boltzmann suppression is mild
enough that the DM particles remain in equilibrium with the SM thermal bath via the
annihilation process $\chi \bar{\chi} \rightarrow e^+ e^-$, governed by the kinetic mixing parameter
$\kappa$. As the universe continues to cool down, this process becomes inefficient and the dark
sector chemically decouples from the SM bath. As discussed in
section~\ref{sec:relicdensity}, the energy density that remains in the dark sector depends
sensitively on the chemical decoupling temperature and hence on $\kappa$. At the same time, the
parameter $\kappa$ determines the dark Higgs boson lifetime, which in the absence of other
interactions can only decay via dark photon loops. In contrast to the case of the coupled
dark sector, the phenomenology therefore depends sensitively on $\kappa$, which should be as
large as possible to ensure that the dark sector is depleted and the dark Higgs bosons
decay sufficiently early.

\begin{figure}[t]
    \centering
    \includegraphics[width=0.48\linewidth]{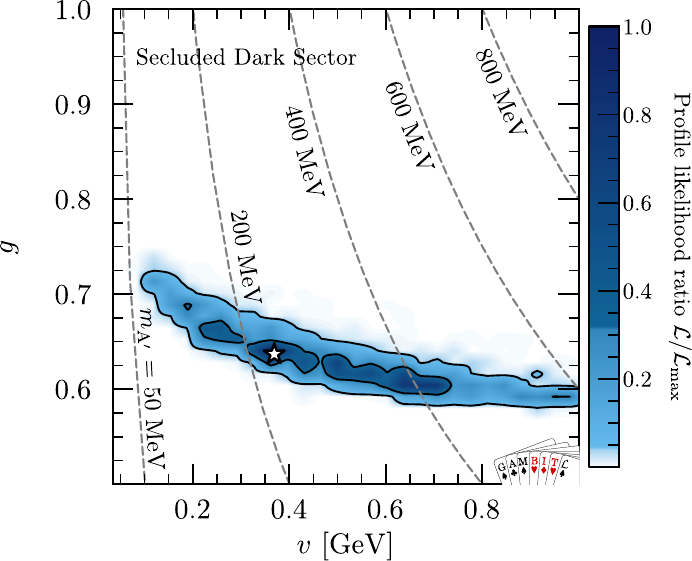}
    \hfill
    \includegraphics[width=0.48\linewidth]{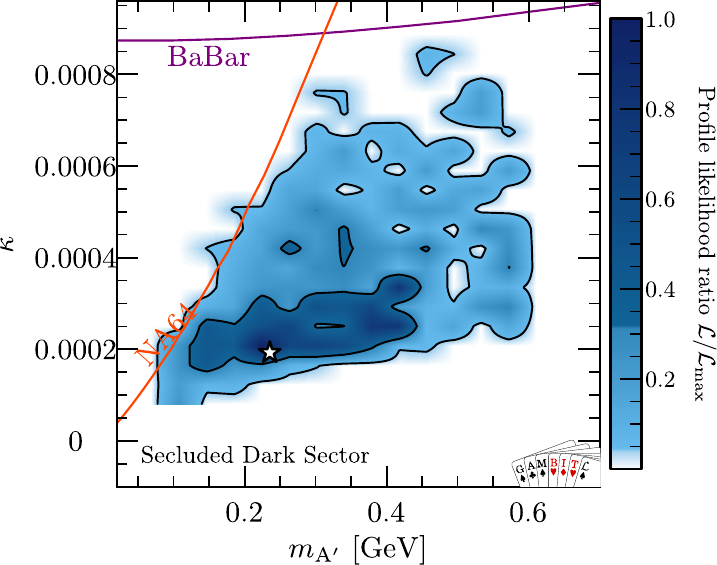}
    
    \includegraphics[width=0.48\linewidth]{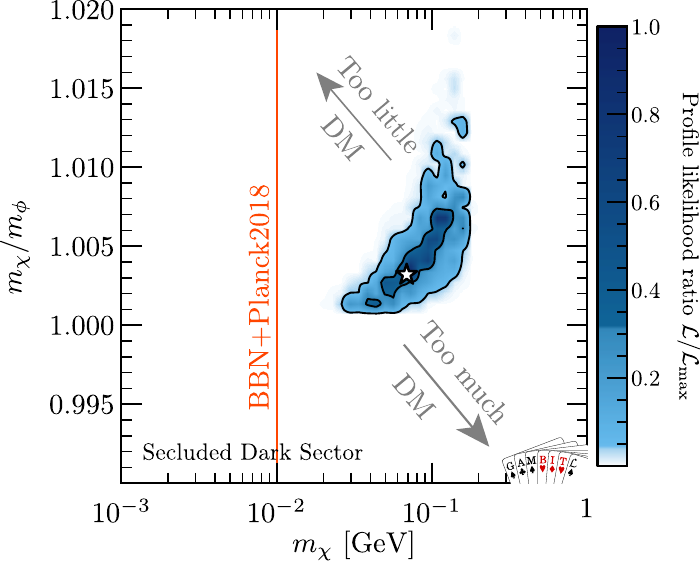}
    \hfill
    \includegraphics[width=0.48\linewidth]{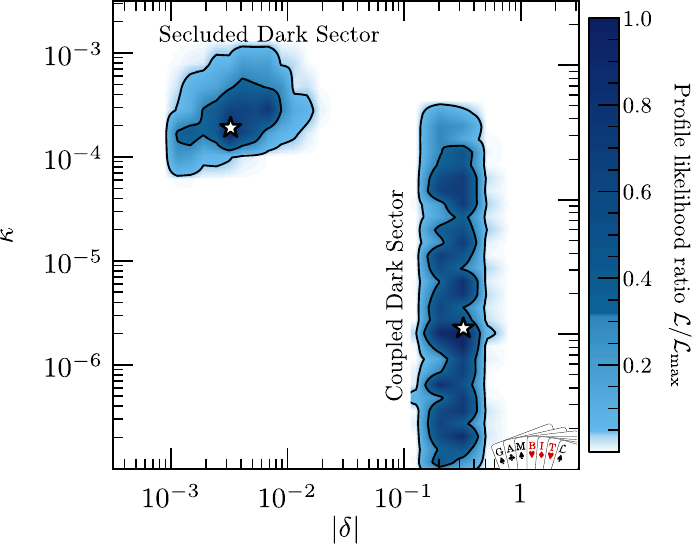}
    
    \caption{Allowed parameter regions determined through a global analysis of the secluded dark sector scenario in terms 
    of the profile likelihood (blue shading). In the top-left panel, dashed lines indicate constant values of the dark photon 
    mass $m_{A'} = g v$.  As in figure~\ref{fig:coupledDS}, we explicitly indicate relevant experimental constraints 
    beyond those from the PTA GW signal. }
    \label{fig:secludedDS}
\end{figure}

We show our main results in figure~\ref{fig:secludedDS}. The top-left panel again shows
that the combination of gauge coupling $g$ and vev $v$ are mainly determined through the
PTA likelihood (cf.~figures~\ref{fig:PTA-GW-likelihood} and \ref{fig:coupledDS}). Note
however, that the best-fit vev $ v \simeq 370 \, \text{MeV}$ is now considerably higher than
the $170 \, \text{MeV}$ found in the equilibrium case. The correspondingly smaller
coupling $g$ means, in comparison, that the phase transition is more strongly supercooled
(larger $\alpha$) and slower (smaller $\beta / H$). This leads to a stronger GW signal, which still
gives a good fit to the PTA data if the peak frequency is also shifted to larger values,
see figure~\ref{fig:pta-data-signal}. We find that the likelihood of the best-fit point is
very similar for the case of a coupled dark sector and a secluded dark sector.

The reason for this shift is explained in the top-right panel of
figure~\ref{fig:secludedDS}, which shows the allowed parameter regions in the
\plane{\kappa}{m_{A'}} plane. We can clearly see that large values of $\kappa$ are in conflict with
the NA64 bound unless increasing the dark photon mass and hence the vev. The best-fit
region corresponds to $m_{A'} \approx 200\text{--}700\,\mathrm{MeV}$ and
$\kappa \approx (1 \text{--} 4) \cdot 10^{-4}$. Correspondingly, also the DM and dark Higgs boson masses
are larger than in the coupled dark sector case. As can be seen in the bottom-left panel
of figure~\ref{fig:secludedDS}, both the dark Higgs boson mass and the DM mass is
typically around $20\text{--}200\,\mathrm{MeV}$ in the allowed parameter
region.

Indeed, as explained in section~\ref{sec:relicdensity}, in the secluded dark sector case
the DM mass must be very slightly heavier than the dark Higgs boson mass, leading to an
allowed parameter region in the range $0.001 \leq \delta \leq 0.01$ in terms of the relative mass
difference $\delta$ defined in eq.~\eqref{eq:delta}. For smaller (or even negative) values of
$\delta$, the DM particles would end up carrying a significant fraction of the energy density
of the dark sector, exceeding the observed DM relic abundance. Larger values of $\delta$, on
the other hand, would lead to a very efficient depletion of DM through annihilations into
dark Higgs bosons and therefore too little DM. We note that the best-fit point corresponds
to a DM Yukawa coupling $y \approx 0.26$, slightly larger than in the coupled dark sector case.
Again, this legitimises {\it a posteriori} our argument that one can approximately ignore
the Yukawa coupling in the effective potential, such that the PTA likelihood mainly
determines $g$ and $v$.

We highlight the difference between the two scenarios in the bottom-right panel of
figure~\ref{fig:secludedDS} in terms of $|\delta|$ and $\kappa$. For the coupled dark sector,
$\kappa$ is largely unconstrained, while $\delta \approx -0.33$. For the secluded dark sector,
$\kappa$ must be as large as possible (without violating bounds from dark photon searches),
while $\delta \approx 10^{-3}$. While the secluded dark sector case requires significant tuning of
the ratio of dark Higgs boson and DM mass, it has the appealing advantage that it makes
testable predictions for laboratory experiments, namely that $\kappa$ should be just below
current exclusion limits. Indeed, these two statements are tightly connected: If the
experimental bound on $\kappa$ were weaker, the dark sector energy density could be depleted
more efficiently before chemical decoupling, and less tuning in $\delta$ would be required to
reproduce the observed DM abundance. The coupled dark sector scenario, on the other hand,
is able to explain both the DM relic abundance and the PTA signal for a wide range of
kinetic mixing parameters at the expense of requiring additional dark Higgs boson decay
channels, which are difficult to probe experimentally (see appendix~\ref{app:toy_model}).

\begin{figure}[t]
    \centering
    \includegraphics[width=0.49\linewidth]{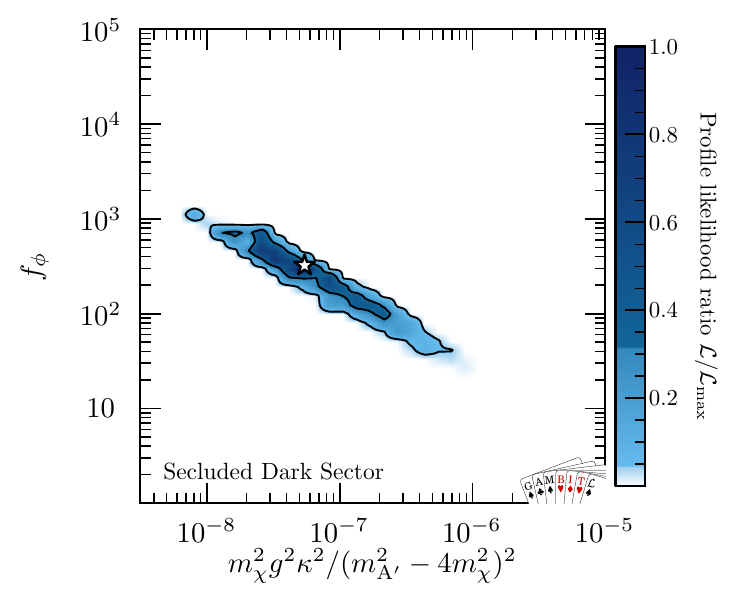}
    \hfill
    \includegraphics[width=0.49\linewidth]{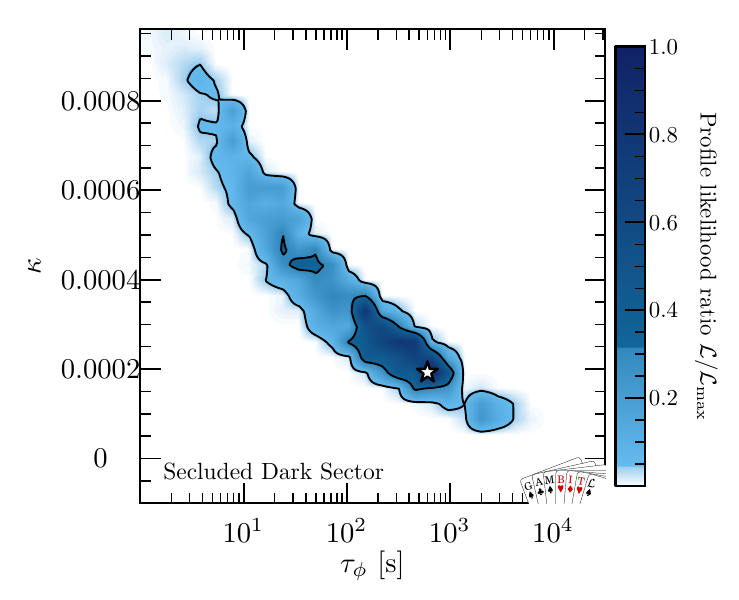}
    
    \includegraphics[width=0.49\linewidth]{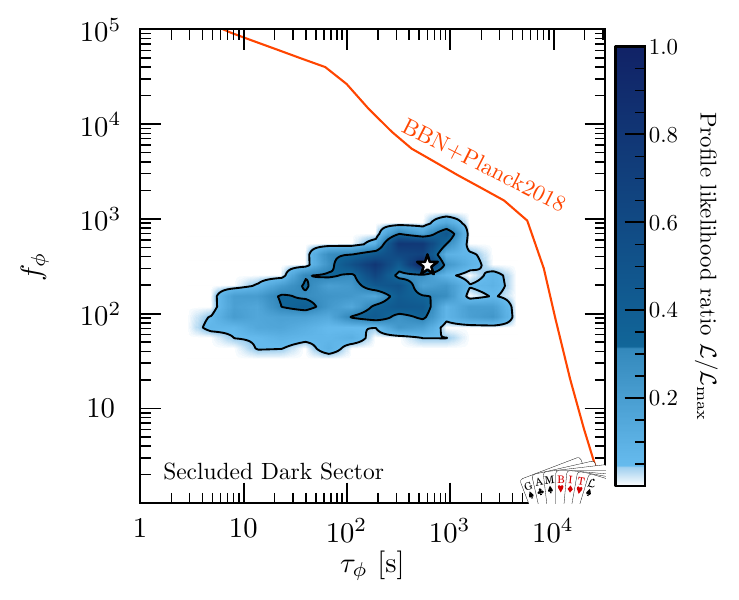}
    \hfill
    \includegraphics[width=0.49\linewidth]{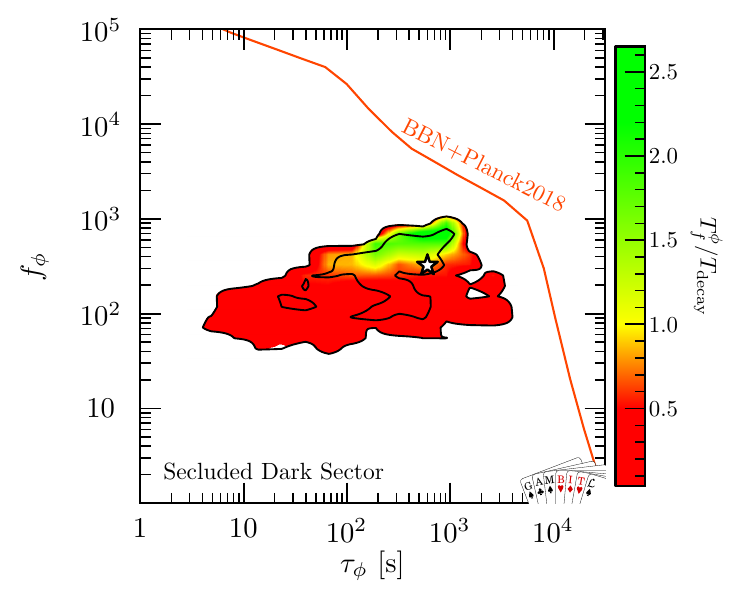}
    
    \caption{Allowed parameter regions for the dark Higgs boson lifetime $\tau_\phi$ and relative would-be abundance 
    $ \fphi = \tilde{\Omega}_\phi / \Omega_\text{DM}$ in terms of the profile likelihood (blue shading) and the largest 
    ratio of 
    decoupling temperature and dark Higgs decay temperature 
    found for a given point in the \plane{\tau_\phi}{f_\phi} plane (red to green shading). The exclusion limit shown in the 
    bottom panels is taken from ref.~\cite{Hufnagel:2018bjp}.}
    \label{fig:secludedDS-decays}
\end{figure}

In figure~\ref{fig:secludedDS-decays}, we take a closer look at the abundance of dark
Higgs bosons and their role in the early universe. The two most relevant parameters in
this context are the dark Higgs boson lifetime $\tau_\phi$ and the relative abundance of dark
Higgs bosons $f_{\phi} = \tilde{\Omega}_\phi/\Omega_{\mathrm{DM}}$ which we parametrise in terms of the
would-be abundance $\tilde{\Omega}_\phi$ of dark Higgs bosons for $\tau_\phi \to \infty$, i.e.\ if the dark
Higgs bosons were to survive to the present day. We emphasise, that this parametrisation
only makes sense if the dark Higgs bosons have chemically decoupled from all other
particle species before they start decaying, an assumption that we will examine shortly.

As argued in section~\ref{sec:relicdensity}, since for $m_\chi > m_\phi$ the dark Higgs bosons
are the dominant component of the dark sector, their abundance is determined by the
chemical decoupling of the dark sector from the SM bath. In the non-relativistic limit,
this decoupling is determined by the parameter combination
$\kappa^2 g^2 m_\chi^2 / (m_{A'}^2 - 4 m_\chi^2)^2$, see appendix~\ref{app:freeze-out}. The top-left
panel of figure~\ref{fig:secludedDS-decays} shows that indeed smaller values of this
parameter correspond to a larger dark Higgs boson abundance.

The dark Higgs lifetime $\tau_{\phi}$, on the other hand, is determined primarily by the kinetic
mixing parameter $\kappa$, which is less precisely determined than $g$ and $v$ and furthermore
enters to the fourth power, see eq.~\eqref{eq:dark_higgs_decay}. In the top-right panel of
figure~\ref{fig:secludedDS-decays} we confirm this expected scaling. We find that the
allowed region of parameter space corresponds to
$10 \, \mathrm{s} \leq \tau_\phi \leq 10^4 \, \mathrm{s}$, implying that the dark Higgs bosons decay
during or after BBN. The lower bound on $\tau_\phi$ is a direct consequence of the upper bound
on $\kappa$ observed already in figure~\ref{fig:secludedDS}.

To understand the upper bound on $\tau_\phi$, we show in the bottom-left panel of
figure~\ref{fig:secludedDS-decays} the allowed parameter regions in terms of the dark
Higgs abundance and lifetime. These are the two parameters that are directly constrained
by cosmological observations, in particular of primordial element abundances and the CMB.
For $\tau_\phi \ll 10^4\,\mathrm{s}$, however, these bounds are actually quite weak compared to
other constraints imposed on the parameter space. For example, at
$\tau_\phi = 100 \, \mathrm{s}$, values as large as $\fphi \sim 10^4$ would be
allowed~\cite{Hufnagel:2018bjp}, while the viable region of parameter space due to other
constraints corresponds to $\fphi < 500$. This picture however changes for
$\tau_\phi \gtrsim 10^4\,\mathrm{s}$, when constraints from photodisintegration become relevant.
Indeed, already for $\tau_\phi = 2\cdot{10}^4 \, \mathrm{s}$ the bound is approximately
$\fphi \lesssim 1$, incompatible with the predictions of our model. Hence BBN constraints
effectively impose $\tau_\phi < 10^4 \, \mathrm{s}$, which translates to the lower bound
$\kappa > 10^{-4}$.

Having determined the relevant region of parameter space, let us briefly examine the
thermal history for the best-fit point of the secluded dark sector scenario, see the right
panel of figure~\ref{fig:history}. The $U(1)^\prime$ breaking succeeds through the nucleation
of bubbles at $2.5 \, \text{MeV}$ which percolate at $1.9 \, \text{MeV}$. Compared to the
coupled dark sector case, we have stronger supercooling and correspondingly larger
reheating. Indeed, after the FOPT the combined bath of SM particles and dark sector
reheats to $32.8 \, \text{MeV}$, which is still well below the masses of the dark sector
states ($m_{A^\prime} = 235 \, \text{MeV}$, $m_{\chi} = 68.6 \, \text{MeV}$,
$m_{\phi} = 68.4 \, \text{MeV}$).
The dark sector decouples chemically from the SM bath at temperatures around
$5.3\,\text{MeV}$, corresponding to
$x_\text{f} \approx 13$. The $\chi\bar{\chi} \to \phi \phi$ conversion processes then remain
efficient until the universe has reached a temperature of around
$T_f^\phi \approx 30 \,\mathrm{keV}$. Decays of the dark Higgs boson start
becoming relevant when $\Gamma_\phi > H(T)$, which is the case for approximately
$T < T_\text{decay} \approx 60 \, \mathrm{keV}$.

At face value, the best-fit point therefore does not satisfy our requirement
$T_f^\phi > T_\text{decay}$, which is needed for the consistency of our relic density
calculation. While it is likely still possible to reproduce the correct DM relic abundance
for some value of $\delta$, finding the allowed parameter regions would require solving
computationally expensive coupled Boltzmann equations. We explore this issue further in
the bottom-right panel of figure~\ref{fig:secludedDS-decays}, where we use colour shading
to show the largest ratio $T_f^\phi / T_\text{decay}$ that we found for a given point in the
\plane{\tau_\phi}{f_\phi} plane, indicating points with $T_f^\phi > T_\text{decay}$
($T_f^\phi < T_\text{decay}$) in green (red) . We observe that
$T_f^\phi \gtrsim T_\text{decay}$ holds for a small region with
$\tau_\phi \gtrsim 10^3 \, \mathrm{s}$, i.e.\ close to the upper bound on $\tau_\phi$ from
photodisintegration. This observation implies that viable parameter regions can be found
that satisfy all requirements simultaneously -- without changing the qualitative picture
presented in figure~\ref{fig:secludedDS} (see appendix~\ref{app:freeze-out} for a more
detailed discussion).

To conclude this discussion, we point out that the energy density of the dark Higgs bosons
is completely negligible when they decay. Even at the most extreme point
($\fphi = 1000$ and $\tau_\phi = 10^3\, \mathrm{s}$), the energy density of
dark Higgs bosons is only around $10^{-2}$ of the radiation energy
density. Hence, the universe does not enter an early period of matter domination and the
amount of entropy injection is completely negligible. For this reason, we do not have to
include a dilution factor for the GW signal and the DM abundance.

\subsection{Secluded dark sector without DM}

In our analysis of the secluded dark sector, we found that the observed DM relic abundance
implies substantial tuning of $\delta$ and a very late decoupling of
$\chi \chi \to \phi \phi$ annihilations. If, instead, we relax the constraint on the abundance of
$\chi$ and only require that it must not exceed the observed DM abundance, much larger
regions of parameter space open up. In particular, $\delta \gg 10^{-3}$ then becomes possible,
such that the DM annihilations into dark Higgs bosons efficiently deplete the DM abundance
before BBN, leaving a decoupled abundance of decaying dark Higgs bosons.

In this case the main constraint on the DM mass comes from the requirement that the Yukawa
coupling $y$ must be small enough to not spoil the strong FOPT. Even though the $\chi$
particles in this case do not account for the dominant form of DM, they play a crucial
role for the cosmological evolution: They facilitate the transfer of the energy density
stored in the Higgs field after the FOPT to the SM via the reaction chain
$\phi \phi \to \chi \chi \to e^+ e^-$ without the need for a dimensionful portal coupling that could spoil
the conformal symmetry of the dark sector at the classical level.

We emphasise that even in this case the role of the fermion $\chi$ is crucial to deplete the
energy density of the dark sector. Without it, the annihilation of dark Higgs bosons into
SM fermions would have to proceed via \emph{two} off-shell dark photons, leading to cross
sections suppressed proportional to $\kappa^4 / m_{A'}^8$. The same is true for the
semi-annihilation process $\phi + e \to 3e$.

\section{Discussion and outlook}
\label{sec:outlook}

The best-fit points obtained in our analyses account for the PTA signal while complying with all available cosmological, 
astrophysical and laboratory constraints. In this section we discuss how the preferred parameter regions can be tested with 
future experiments and GW observations and emphasise the importance of improved theory predictions.

The upcoming IPTA data release will combine the data sets collected by the individual collaborations and is expected to 
establish decisive evidence in favor of a GW background, i.e.\ a detection of of the Hellings-Downs correlation with 
significance above $5 \sigma$~\cite{NANOGrav:2020spf}. Moreover, this data release may provide new information on the 
anisotropy of the GW amplitude over the celestial sphere (see ref.~\cite{Grunthal:2024sor} for recent results by the Meerkat 
collaboration) or on individual sources of continuous GW signals~\cite{IPTA:2023ero, Antoniadis:2023aac, Agazie:2024jbf}, 
both of which would favour an astrophysical origin of the PTA signal. While the current status is inconclusive, any future 
update should be included into 
global fits in order to study the impact on the preference for or against a FOPT 
interpretation of the signal.

At the same time, improved constraints on the GW spectrum could shed further light on its possible origin. In the coupled 
dark sector case, the peak frequency and amplitude of the GW signal can vary significantly, so that we expect the model to 
provide a good fit also to future measurements. In the secluded dark sector case, on the other hand, the vev has to be quite 
large in order for the model to satisfy cosmological constraints on out-of-equilibrium dark Higgs boson decays. This leads to 
the clear prediction that the GW signal extends beyond the currently measured frequency range, see 
figure~\ref{fig:pta-data-signal}. If such a rise is not seen in future data~\cite{Janssen:2014dka, Sesana:2019vho}, this would 
disfavour the secluded dark sector case. 

\begin{figure}[t]
\centering
\includegraphics[width=0.49\linewidth]{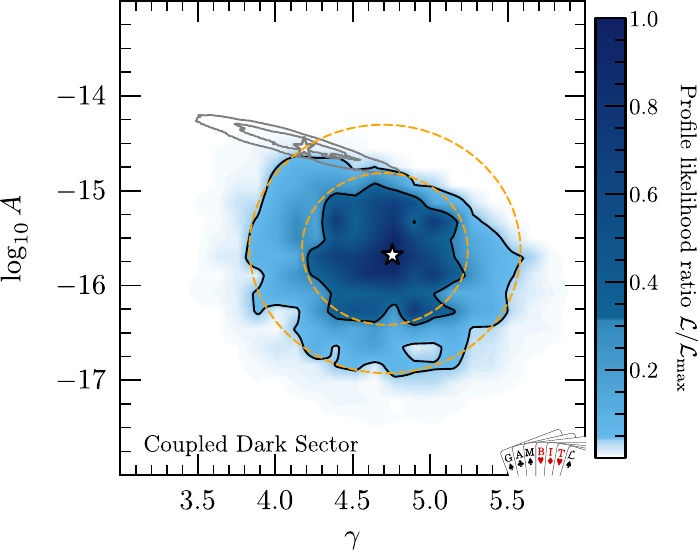}
\caption{Allowed parameter regions for the two parameters characterising the slope ($\gamma$) and amplitude ($A$)
    of the SMBH 
    contribution to the GW signal (see section~\ref{sec:GWSMBHB}). The yellow dashed lines indicate the allowed 
    parameter regions without considering PTA data, i.e.\ based on expectations from astrophysical models and simulations, 
    cf.~eq.~(\ref{eq:prior-smbh}). The blue shaded regions correspond to the allowed parameter regions 
    found in our analysis. For comparison we show with grey contours the preferred values and best-fit point in the case that 
    the PTA signal is fitted by SMBH binaries alone, still using the likelihood from eq.~\eqref{eq:prior-smbh} but setting the FOPT contribution to zero.} 
    \label{fig:SMBH}
\end{figure}

Furthermore, it will be essential to improve our understanding of the SMBH binary contribution to the GW spectrum. In the 
present work, we have used a rather simple likelihood for the amplitude $A$ and slope $\gamma$ of the signal, see 
section~\ref{sec:GWSMBHB}. This choice of likelihood also affects the best-fit parameter regions for $A$ and $\gamma$ 
found in our scans. This is illustrated in figure~\ref{fig:SMBH}, which shows the allowed parameter regions for the two parameters without PTA likelihood as given in eq.~(\ref{eq:prior-smbh}) (yellow ellipses), when fitting the PTA likelihood with only an SMBH contribution and no dark sector (grey 
contours) 
and when including also a FOPT contribution (blue shading).
In the latter case we consider the coupled dark 
sector case for concreteness, but the corresponding plot for the secluded dark sector case looks almost identical. When trying to fit the PTA data with an SMBH contribution alone, we find a significant shift in the allowed parameter region relative to the astrophysical expectation, indicating the underlying tension. When including a FOPT contribution, on the other hand, the only impact of the PTA data on the preferred parameter regions is to disfavour values of $A$ and $\gamma$ that would 
imply a too large GW signal (upper right corner).

An interpretation of the observed GW signal purely in terms of SMBH binaries without contribution from a FOPT is 
disfavoured by the combined constraint on $A$ and $\gamma$ that we impose, cf.~eq.~\eqref{eq:prior-smbh}, motivated by 
astrophysical models of  SMBH populations~\cite{NANOGrav:2023hvm}. The NANOGrav collaboration has shown 
that the SMBH signal can only account for a fraction of the observed GW signal when assuming typical population 
properties~\cite{NANOGrav:2023hfp}. Recently, also $N$-body simulations including an on-the-fly treatment of dynamical 
friction on SMBH mergers have been performed, coming to the conclusion that there might be a need for either a yet 
unexplained black hole-host galaxy mass relationship or alternatively new physics~\cite{Chen:2025wel}. Intriguingly, our 
best-fit GW spectra (see figure~\ref{fig:pta-data-signal}) match the missing power needed to explain the observed, 
unexpectedly high GW amplitudes at frequencies above $3 \, \text{nHz}$ -- which SMBHs cannot account for due to their 
scarcity. 

\begin{figure}[t]
    \centering
    \includegraphics[width=0.5\linewidth]{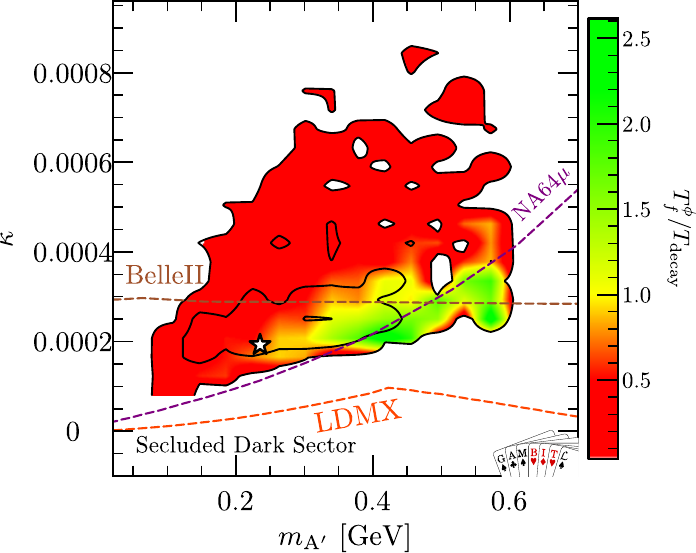}
    \caption{Projected sensitivities of searches for invisibly decaying photons at Belle II, as well as for the proposed LDMX 
    and NA64$\mu$ experiments in the \plane{m_{\mathrm{A}^\prime}}{\kappa} parameter plane of the secluded dark sector 
    model. The colour shading indicates the ratio $T_f^\phi / T_\text{decay}$; ratios smaller than unity indicate that
    our relic density calculation may be unreliable.} \label{fig:sds_sensitivities}
\end{figure}

The model that we consider can also be further constrained by future searches for invisibly decaying dark photons at 
Belle II~\cite{Belle-II:2018jsg} and fixed-target experiments, see figure~\ref{fig:sds_sensitivities}. Upcoming runs of the NA64 experiment using muon 
beams~\cite{Gninenko:2653581}, as well as the proposed fixed-target experiment LDMX~\cite{LDMX:2018cma} will 
significantly improve the bound on the kinetic mixing parameter $\kappa$ for sub-GeV dark photons. In fact, for the 
secluded dark sector, the entire allowed range of $\kappa$ found in our analysis will be probed by LDMX. A null result 
would hence rule out the case of a secluded dark sector and instead prefer the coupled dark sector case, where $\kappa$ 
may be much smaller. In this case, the most promising avenue is to study the implications of the assumed effective coupling 
of the dark Higgs boson to electrons~\cite{BaBar:2020jma,Cogollo:2024fmq}, for example through probes of lepton-flavor violation (see 
appendix~\ref{app:toy_model} and recent work on leptophilic axion-like particles~\cite{Chun:2021rtk,Altmannshofer:2022ckw,Armando:2023zwz,Liu:2023bby}).

Also future cosmological probes may provide valuable additional information for determining the origin of the nHz GW 
signal. Stronger bounds on $\Delta N_\text{eff}$ from CMB-S4~\cite{CMB-S4:2022ght} would place an upper bound on the 
amplitude of the GW background itself and tighten  bounds on changes in the neutrino temperature and the baryon-to-
photon ratio due to DM annihilations and dark Higgs boson decays. However, even if CMB-S4 confirms an SM value of 
$N_\text{eff}$, this could be accommodated in our model for a large enough vev, corresponding to large DM mass. Finally, 
future probes of the spectral shape of the CMB black-body spectrum could give relevant constraints on the interpretation of 
the PTA signal, as recently argued in ref.~\cite{Ramberg:2022irf}.

At the same time, further insights into the models under consideration may also come from theoretical developments. Our 
calculation of the bubble nucleation rate, for example, is subject to considerable uncertainties and depends on arbitrary 
gauge choices. Over the past decade, a rigorous effective field theory approach for the resummation of infrared-divergent 
modes has
been developed, allowing for a more accurate computation of phase transition properties
than the daisy-improved potential that we have implemented~\cite{Croon:2020cgk, Ekstedt:2022bff}. This so-called 
dimensional reduction approach has recently been shown to also been applicable to the case of strong
supercooling~\cite{Kierkla:2023von}, even though it is based on a high-temperature expansion.
Nevertheless, it was shown in ref.~\cite{Lewicki:2024xan} that using dimensional reduction instead of the conventional 
calculation only leads to negligible shifts in the inferred regions of allowed parameter space.

Furthermore, we also anticipate methodological advances in the description of strong 
supercooling~\cite{Ellis:2020nnr, Gouttenoire:2023naa, Lewicki:2023ioy, Lewicki:2024sfw}, the determination of the bubble 
wall velocity~\cite{Ekstedt:2024fyq} and the extraction of GW spectra from numerical simulations of phase transitions with 
large and fast bubbles including turbulence~\cite{Caprini:2024gyk, Dahl:2024eup}. A better understanding of these effects 
may lead to shifts in the preferred region of parameter space, even in the absence of new data. Alternatively, in the absence 
of a more accurate description, it could be interesting to include additional nuisance parameters in our analysis to 
parametrise theoretical uncertainties and study how these translate into larger allowed parameter regions. A similar 
comment applies to the relic density calculation: A detailed treatment of the coupled Boltzmann equations for dark Higgs 
bosons and DM, fully taking into account the effects of kinetic decoupling~\cite{Binder:2021bmg, Aboubrahim:2023yag} and 
number-changing processes~\cite{Farina:2016llk,Cline:2017tka,Bringmann:2020mgx} 
would be desirable in order to accurately determine the 
preferred values of the DM mass in our set-up.

\section{Conclusions}
\label{sec:conclusions}

The stochastic gravitational wave background measured by various pulsar timing arrays (PTAs) such as NANOGrav offers 
a tantalising glimpse of unknown astrophysics and the very early universe.
In this work we explored the possibility that the gravitational wave signal arises dominantly from a first-order phase 
transition in a classically conformal dark sector. We demonstrated that the phase transition in the dark sector -- consisting of 
a dark Higgs boson, a dark photon and a fermionic dark matter candidate charged under a new $U^{\prime}(1)$ symmetry 
-- can be strong enough to fit the PTA data,  while simultaneously giving mass to the dark matter particle, which 
subsequently obtains its relic abundance through interactions with dark photons and dark Higgs bosons.

The nano-Hertz frequencies and the peak amplitude of order
$h^{2}\Omega_{\mathrm{GW}} \sim \mathcal{O}(10^{-8})$ of the PTA data require the mass scale of the dark
sector particles to be around $\mathcal{O}(100\,\mathrm{MeV})$ and the phase transition to be
strongly supercooled, with transition temperatures around
$T \sim \mathcal{O}(1\,\mathrm{MeV})$. Therefore a significant amount of energy is
present in the dark sector after the phase transition, which can impact the subsequent
cosmological evolution if it is not efficiently transferred to the Standard Model before
the beginning of Big Bang Nucleosynthesis (BBN).

We have studied two ways in which the energy density of the dark sector is depleted after the phase transition,
which we dubbed \textit{coupled} and \textit{secluded dark sector}, respectively. For the coupled scenario we assumed efficient decays of the dark 
Higgs boson into electrons, arising from higher-dimensional operators. In the secluded scenario, the energy density is 
transferred to the SM through a kinetic mixing of the dark and SM photon. Laboratory constraints on the kinetic mixing 
parameter however make it impossible to fully deplete the abundance of dark Higgs bosons before BBN, so that a sizable 
population of particles remains that decay out of equilibrium via loop-induced processes. These decays can impact various 
cosmological observables, such as the number of relativistic degrees of freedom and the primordial element abundances.

We performed global fits for both scenarios including all relevant cosmological and laboratory constraints,
identifying viable regions of parameter space
that can fit the PTA data and reproduce the dark matter (DM) relic density. We find that in the coupled 
dark sector case the vev and the dark gauge coupling are dominantly constrained by the PTA likelihood, while the DM mass 
is constrained by the relic density requirement. 
Since the dark sector remains in chemical equilibrium
through the effective electron coupling, the kinetic mixing parameter remains
unconstrained. This also ensures that the dark sector particles are Boltzmann suppressed
during BBN such that the standard cosmological history is recovered.

On the other hand, the secluded dark sector -- without the effective electron coupling --
requires as large kinetic mixing as possible, i.e.~as allowed by laboratory searches for dark photons, to deplete the dark sector energy density through annihilations of the dark 
matter particles into electrons before the dark sector decouples. After decoupling from the Standard Model, the DM 
particles and dark Higgs bosons continue to annihilate into each other. To reproduce the observed DM relic 
abundance, it is necessary for the DM particles to be slightly heavier than the dark Higgs bosons (at the 
permille-level), such that most of the DM particles annihilate into dark Higgs bosons, which subsequently 
decay into electrons. 
The strongest constraints on this scenario derive  from the potential 
photodisintegration of light nuclei, which implies a dark Higgs boson 
lifetime $\tau_{\phi} < 10^{4}\,$s. In view of current 
laboratory constraints on the kinetic mixing parameter, the only way to achieve such small 
lifetimes is through dark Higgs boson masses that are somewhat larger than in the coupled sector case. 
The decoupled sector scenario therefore also requires a larger peak amplitude, 
in order to maintain  a good fit to the PTA data,
and hence stronger supercooling. 

Future searches for dark photons such as Belle II and NA64$\mu$ will 
provide the 
possibility to fully test the secluded sector
scenario by searching for dark photons. For the coupled sector case, on the other hand, studying the 
implications of (potentially flavour-violating)
coupling of dark Higgs boson to electrons will be more relevant~\cite{BaBar:2020jma,Cogollo:2024fmq}.
At the same time, the upcoming third IPTA data release is expected to improve our understanding of the nature and the exact spectral shape of the nano-Hertz 
gravitational wave signal. This will be an important step towards understanding its origin, and in particular
to disentangle the leading explanations in terms of merging supermassive black holes or dark sector phase transitions -- or a combination thereof,
as studied in this work.  

\hfill

\acknowledgments

We thank Csaba Balazs, Esau Cervantes, Andrew Fowlie, Will Handley, Jean Kimus, Thomas Konstandin, Marek Lewicki, V.M.~Sabarish, Kai Schmidt-Hoberg, Michel Tytgat and Miguel Vanvlasselaer for discussions, Jorinde van de Vis for help with \textsf{WallGo} and Peter Athron, Tom\'{a}s E.\ Gonzalo, J\"{o}rn Kersten, Anders Kvellestad and Pedro Schwaller for valuable comments on the manuscript. SB acknowledges the support by the Doctoral School `Karlsruhe School of
Elementary and Astroparticle Physics: Science and Technology'. TB gratefully acknowledges
support through a FRIPRO grant of the Norwegian Research Council (project ID 353561
`DarkTurns'). FK acknowledges funding by the Deutsche Forschungsgemeinschaft (DFG) through
Grant No. 396021762 -- TRR 257. CT acknowledges funding through Germany's Excellence Strategy EXC 2121 ``Quantum Universe'' Grant No.~390833306. TB and FK are grateful to the Mainz Institute for
Theoretical Physics (MITP) of the Cluster of Excellence PRISMA+ (Project ID 390831469),
for its hospitality and its partial support during the completion of this work. This work
was performed using the Cambridge Service for Data Driven Discovery (CSD3), part of which
is operated by the University of Cambridge Research Computing on behalf of the STFC DiRAC
HPC Facility (\url{www.dirac.ac.uk}). The DiRAC component of CSD3 was funded by BEIS
capital funding via STFC capital grants ST/P002307/1 and ST/R002452/1 and STFC operations
grant ST/R00689X/1. DiRAC is part of the National e-Infrastructure. The plots were made
using pippi v2.1~\cite{Scott:2012qh}.

\appendix

\section{A simple model for effective electron couplings}
\label{app:toy_model}

Effective couplings to electrons can be generated by introducing two heavy
vector-like leptons $E$ and $L$ that carry the same charge under the SM gauge group as the right-handed electron $e_R$ 
and the left-handed lepton doublet $\ell_L$, respectively, as well as the same charge under the $U(1)'$ as the dark Higgs 
boson. The Lagrangian for these new fermions, excluding the gauge interactions, is then given by
\begin{equation}
    \mathcal{L} \supset m_E \bar{E}_L E_R + m_L \bar{L}_L L_R + y_R \phi \bar{E}_L e_R + y_L \phi^\dagger \bar{\ell}_L L_R + y_{LE} \bar{L}_L E_R H + y_{EL} \bar{E}_L L_R H^\dagger + \text{h.c.}
\end{equation}
If the masses $m_{E,L}$ are sufficiently large, the vector-like fermions can be integrated out to give the effective interaction
\begin{equation}
\label{eq:eff_op}
    \frac{1}{\Lambda^2} \phi^\dagger \phi \bar{\ell}_L e_R H + \text{h.c.}
\end{equation}
Once the electroweak symmetry and the $U(1)'$ symmetry are broken, one then obtains the effective interaction 
$y_\text{eff} \phi \bar{e} e$ with $y_\text{eff} = v_h v_\phi / (2 \Lambda^2)$. This interaction gives rise to sufficiently fast 
decays of the dark Higgs boson, provided that $\Lambda < 100$ -- 1000 TeV.

If the vector-like leptons couple to all lepton families, they would generate the decay $\mu \to e \phi$, which is tightly 
constrained by experiments~\cite{Cornella:2019uxs}, implying $\Lambda > 1000$ TeV if the dark Higgs boson is long-lived 
and even stronger bounds on $\Lambda$ if the dark Higgs decays promptly into an $e^+ e^-$ pair. We therefore need to 
assume that the vector-like leptons couple dominantly to the first generation, in which case experimental bounds on 
$\Lambda$ are significantly relaxed.

The effective coupling of the dark Higgs to electrons can also be constrained by observations of supernova SN1987A, in 
analogy to the case of axion-like particles~\cite{Carenza:2021pcm}. However, to accurately calculate the cooling due to 
dark Higgs boson emission, it is necessary to consider also the production of DM particles and dark photons via kinetic 
mixing~\cite{Sung:2021swd}, which contribute to the trapping of dark Higgs bosons. To complicate matters further, the 
temperature inside the supernova may be sufficient to restore the symmetry in the dark sector, such that the decay 
$\phi \to e^+ e^-$ would be forbidden in the supernova core. A detailed study of these interesting effects is beyond the 
scope of the present work.

Finally, we note that the explicit mass terms for the vector-like leptons break the conformal symmetry in the dark sector. 
Loops involving these heavy fermions may therefore reintroduce a mass term for the dark Higgs boson. Moreover, 
there may be loop contributions to both kinetic mixing and Higgs mixing. 
A detailed study of how the dark sector that we 
consider may arise from possible UV completions is left for future work. 

\section{Estimating the bubble wall velocity}
\label{app:darkwalls}

The bubble wall velocity depends on the friction exerted on the bubble walls by the particles that become massive as they 
cross from the false to the true vacuum. Since our model also includes a gauge boson $A^\prime$, which obtains its mass 
in the phase transition, splitting radiation gives rise to additional friction in the bubble walls~\cite{Hoche:2020ysm,
Gouttenoire:2021kjv}, which grows with the Lorentz factor $\gamma = 1/\sqrt{1 - v_\text{w}^2}$ of the bubble wall. This 
friction term prevents the walls from running away, i.e.\ they cannot accelerate to ultra-relativistic velocities before collision.

\begin{figure}[]
  \centering
  \includegraphics[width=0.6\textwidth]{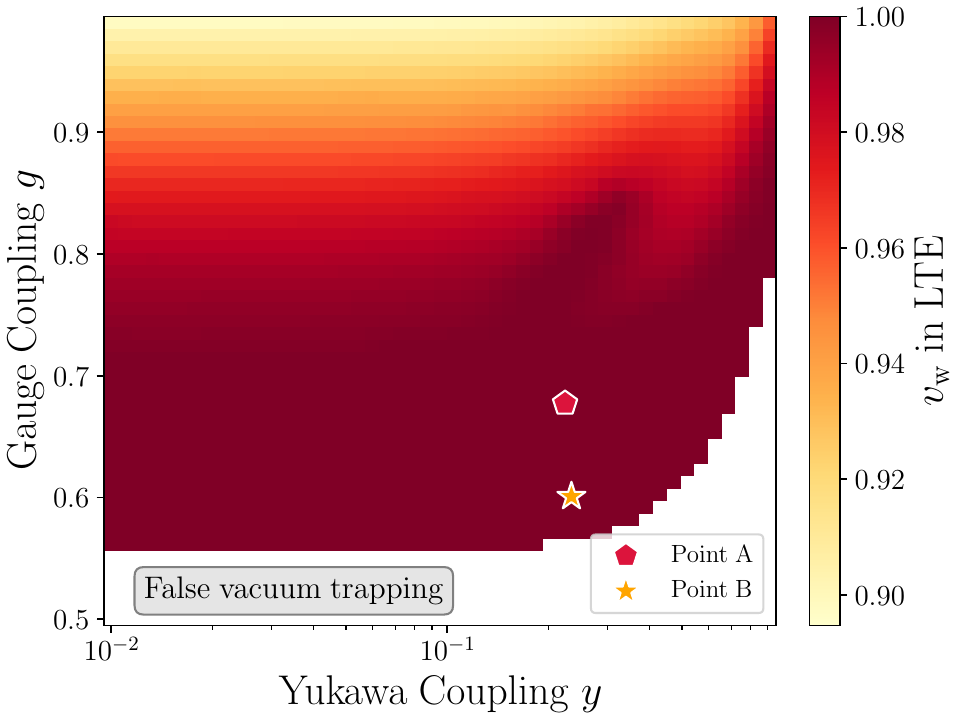}
  \caption{Bubble wall 
  velocities calculated in the LTE approximation as a function of $g$ and $y$. The coupling values of the parameter points A  
  and B from table~\ref{tab:bestfit} are indicated by the red pentagon and the yellow star, respectively. For concreteness, we 
  have set $v = 100\,$MeV, but the result is almost completely independent of the vev.}
  \label{fig:wall-velo}
\end{figure}

A more detailed computation of the bubble wall velocity is quite challenging, since
model-dependent out-of-equilibrium effects have to be taken into account. However, recently it was shown that for 
deflagration and hybrid solutions, assuming local thermal equilibrium (LTE) gives a good approximation to full 
model-dependent calculations of the wall velocity
\cite{Laurent:2022jrs}. Here we make use of the LTE approximation, using the code provided in ref.~\cite{Ai:2023see} 
for the calculation, and illustrate the resulting wall velocity in figure~\ref{fig:wall-velo}.

For detonations, however, there exist no stable solutions in the LTE approximation, since
the pressure is a monotonically decreasing function of $v_\text{w}$~\cite{Ai:2023see}. 
We thus employ the recently published code \textsf{WallGo}
\cite{Ekstedt:2024fyq} to compute out-of-equilibrium effects for the 
wall velocity. Implementing our model in \textsf{WallGo}, we 
verified that around the best-fit points of our scans the inclusion of out-of-equilibrium 
effects only results in a sub-percent level deviation from the LTE approximation of the bubble wall velocity
shown in figure~\ref{fig:wall-velo}.

We conclude that the bubble walls in our model 
are relativistic but not ultra-relativistic when they collide, such that sound waves are 
indeed expected to give the dominant contribution to the GW signal, as assumed in eq.~\eqref{eq:GWspec}.

\section{Dark matter freeze-out in a secluded dark sector}

\label{app:freeze-out}

Here we describe in detail our treatment of a secluded dark sector as described in section 
\ref{sec:dec_dark_sector} (see also figure \ref{fig:thermalisation-diagrams}). In this situation, the `hidden'
interactions $\chi\chi\leftrightarrow \phi\phi$ are more efficient than annihilation and creation processes
$\chi\chi\leftrightarrow e^+e^-$ that keep the DM particles in chemical equilibrium with the SM. In other words,
those two processes freeze out at different temperatures $T_f^\phi$ and $T_f^\text{SM}$, respectively, 
with $T_f^\phi<T_f^\text{SM}$. As explained in the following, this allows us to  obtain the final relic density
of $\chi$ and $\phi$ in two steps.

In the first step, we compute $T_f^\text{SM}$ along with the {\it would-be} DM relic density $\tilde \Omega_\chi h^2$ if one 
were to completely ignore the $\chi\chi\leftrightarrow \phi\phi$ processes, using the standard numerical routines 
available in \ds. We note that until chemical decoupling from the SM, $T\lesssim T_f^\text{SM}$, {both} $\chi$ 
and $\phi$ are in full chemical equilibrium and hence follow Maxwell-Boltzmann distributions. 
At these temperatures, the ratio of their abundances is thus given by
\begin{equation}
\label{Yratioxpsi}
\frac{Y_\chi}{Y_\phi}=\frac{Y_\chi^{\rm eq}}{Y_\phi^{\rm eq}}= 2\, %\frac{g_\chi}{g_\phi}
(1-\delta)^{-3/2} e^{-\delta x}\,,
\end{equation}
where $x\equiv m_\chi/T$ and $\delta\equiv1-m_\phi/m_\chi$
as introduced in eq.~(\ref{eq:delta}).
Here, the leading factor of 2 takes into account that the DM particles have 2 spin d.o.f.~while
the dark Higgs bosons only have 1.

In the second step, we estimate by how much the relic abundance $\tilde \Omega_\chi h^2$ is changed due to the
$\chi\chi\leftrightarrow \phi\phi$ processes happening after freeze-out from the SM, 
during $T_f^\text{SM}>T>T_f^\phi$. To do so, we 
observe that detailed balance in this regime implies  the following  two conditions:
\begin{align}
Y_\chi + \tfrac{1}{2} Y_\phi &= \text{const}\,, \label{eqapp:constant_number} \\
 Y_\chi^2 \langle \sigma v \rangle_{\chi\chi\to\phi\phi} &= Y_\phi^2 \langle \sigma v \rangle_{\phi \phi \to \chi \chi} \, ,
 \label{eqapp:constant_rate}
\end{align}
where the factor 1/2 arises because $\phi$ is its own antiparticle. 
Note that the first equation assumes that dark Higgs boson decays are not (yet) efficient for $T>T_f^\phi$. 
We can evaluate the above two  conditions at both $T_f^\text{SM}$ and $T_f^\phi$ to find the final DM relic abundance
$\Omega_\chi h^2$, making use of eq.~(\ref{Yratioxpsi}) at $T=T_f^\text{SM}$ as well as 
$\Omega_\chi h^2/\tilde \Omega_\chi h^2 = Y_\chi(T_f^\phi) / Y_\chi(T_f^\text{SM})$. We thus arrive
at our central result
\begin{equation}
\label{eq:oh2final}
\Omega_\chi h^2 \simeq {\tilde \Omega_\chi h^2}\times
\frac{
1+\frac12\frac{g_\phi}{g_\chi}(1-\delta)^{3/2}e^{\delta x_f^\textrm{SM}}
}
{1+\frac12\sqrt{\overline r(x_f^\phi)}}\,,
\end{equation}
where we introduced
\begin{equation}
\overline r\equiv \frac{ \left \langle \sigma v\right\rangle_{\chi\chi\to\phi\phi}}{ \left \langle \sigma v\right\rangle_{\phi\phi\to\chi\chi}}
\end{equation}
as a measure of how `forbidden' one direction of the reaction is with respect to the other.
Neglecting decay, the abundance of the dark Higgs bosons follows in full analogy as  
\begin{equation}
Y_\phi =\left (Y_\chi\right)_{x_f^\textrm{SM}}\left(1+\frac12  \frac{g_\phi}{g_\chi}(1-\delta)^{3/2}e^{\delta x_f^\textrm{SM}}\right)
\left(\frac{\sqrt{\overline r}}{1+\frac12  \sqrt{\overline r}} \right)_{x_f^\phi} \,.
\end{equation}
While $\Omega_\chi h^2$ should be compared to the observed DM abundance of $\Omega_\textrm{DM} h^2= 2\Omega_\chi h^2 =0.12$~\cite{Planck:2018vyg}, a primordial abundance of decaying dark Higgs bosons is subject to the
constraints discussed in section \ref{sec:cosmo_constraints}.

The missing piece left to evaluate in our central result for the relic density, eq.~(\ref{eq:oh2final}), is 
the quantity ${\overline r(x_f^\phi)}$. For this, we first note that
\begin{equation}
 \frac{ \left ( \sigma v\right)_{\chi\chi\to\phi\phi}}{ \left (\sigma v\right)_{\phi\phi\to\chi\chi}}
 =\frac12\frac14\frac{\sqrt{s-4m_\phi^2}}{\sqrt{s-4m_\chi^2}}\,,
%  this ratio just comes from the phase space, i.e. independent of $t$ !
\end{equation}
where the first factor ($1/2$) takes into account that only the $\phi$ particles are
self-conjugate, while the second factor ($1/4$) arises due to the different spin d.o.f..
Using the standard expressions for the thermally averaged cross section~\cite{Gondolo:1990dk}, we then
obtain
\begin{align}
\overline r\equiv \frac{ \left \langle \sigma v\right\rangle_{\chi\chi\to\phi\phi}}{ \left \langle \sigma v\right\rangle_{\phi\phi\to\chi\chi}}&= 
\frac{m_\chi\, {K_2}^2(m_\phi/T)}{m_\phi\, {K_2}^2(m_\chi/T)}\frac{\int d\tilde s\,{ \left ( \sigma v\right)_{\chi\chi\to\phi\phi}} \sqrt{\tilde s-1}(2\tilde s-1)K_1\left(\frac{2\sqrt{\tilde s}m_\chi}{T}\right)}{\int d {\tilde s_\phi}\,{ \left (\sigma v\right)_{\phi\phi\to\chi\chi}}\sqrt{ {\tilde s_\phi}-1}(2 {\tilde s_\phi}-1)K_1\left(\frac{2\sqrt{ {\tilde s_\phi}}m_\phi}{T}\right)}\\
&\approx
\frac{(1-\delta)^3 {K_2}^2[x(1-\delta)]}{{8 K_2}^2[x]}
\frac{\int d\tilde s\,\sqrt{\tilde s-1}(2\tilde s-1)K_1\left[2\sqrt{\tilde s}x\right]}
{\int d\tilde s\, \sqrt{\tilde s-1}\left(2 \tilde s(1-\delta)^{-2}-1\right)K_1\left[2\sqrt{\tilde s}x\right]} \,, \label{eq:rbarfinal}
\end{align}
where $\tilde s\equiv s/(4m_\chi^2)$,
{$\tilde s_\phi\equiv s/(4m_\phi^2) = \tilde s (1-\delta)^{-2}$} and all integrations are restricted to
$\tilde s\geq \max(1,(1-\delta)^2)$. The last step holds for $s$-wave,
$\sigma v\approx const.$, but is straight-forward to generalise for other partial waves. In our
numerical implementation we update \ds\ to evaluate eq.~(\ref{eq:rbarfinal}) at dark sector
freeze-out, which we estimate from the usual freeze-out condition
$n_\chi^2 \left \langle \sigma v\right\rangle_{\chi\chi\to\phi\phi} = n_\phi^2 \left \langle \sigma v\right\rangle_{\phi\phi\to\chi\chi} \sim n_\chi H$. Using the
above conditions, in other words, we numerically solve
\begin{equation}
\label{eq:Tdecphi}
Y_\chi (x_f^{\rm SM})\frac{1+\frac12\frac{g_\phi}{g_\chi}(1-\delta)^{3/2}e^{\delta x_f^{\rm SM}} }{1+\frac12\sqrt{\overline r(x_f^\phi)}}
 \left \langle \sigma v\right\rangle_{\chi\chi\to\phi\phi} (x_f^\phi) = H(x_f^\phi) / s(x_f^\phi)
\end{equation}
to obtain $x_f^\phi = m_\chi/T_f^\phi$. 

\begin{figure}[t]
  \centering
  \includegraphics[width=0.6\textwidth]{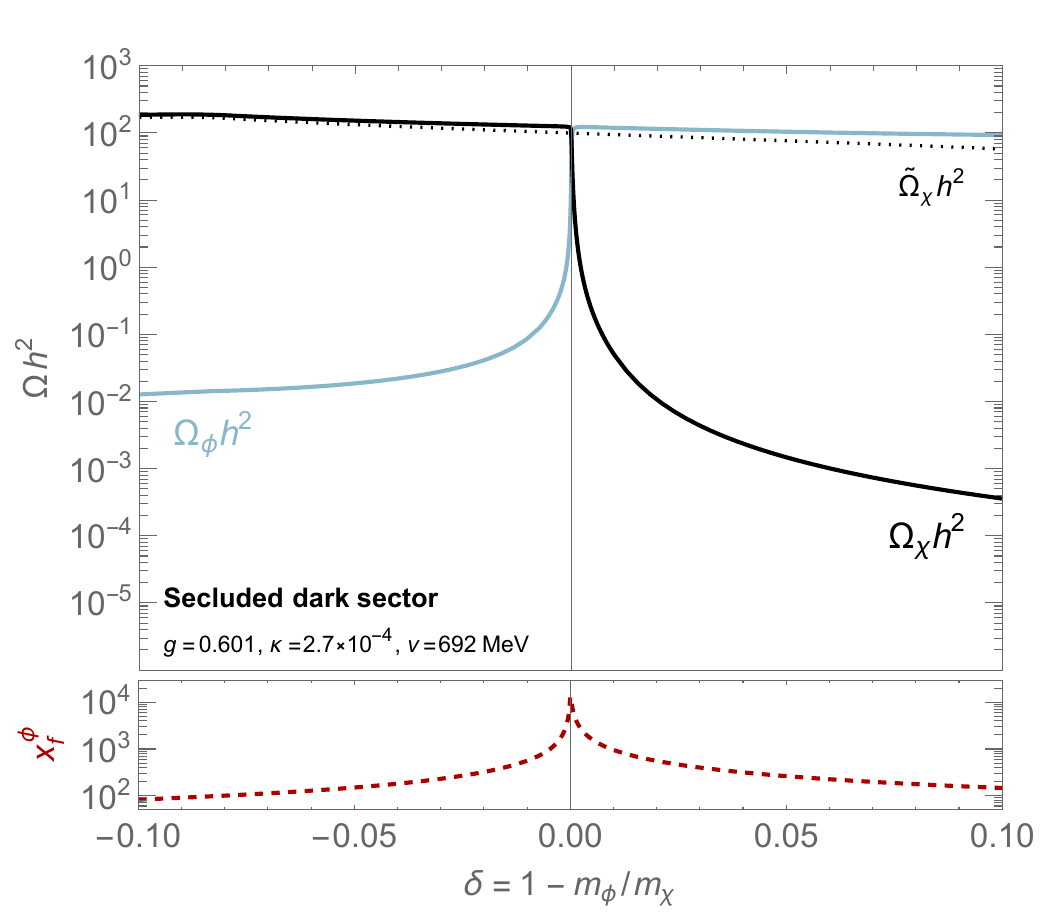}
  \vspace{-2mm}
  \caption{{\it Top.} DM relic density (solid black line) 
   in the secluded dark sector scenario as a function of the mass splitting $\delta$, with the other model 
  parameters set to the values of Point B in table~\ref{tab:bestfit}. The corresponding
  would-be relic density of the scalar particles $\phi$, neglecting their eventual decay, is given  by
  the solid blue line. 
  The dotted black line shows the DM relic density that would result if one were to 
  neglect the $\chi\chi\leftrightarrow\phi\phi$ interactions. {\it Bottom.} Value of $x=m_\chi/T$
  indicating the freeze-out of these number-changing dark-sector interactions. 
  All curves are computed using \ds.
  } 
  \vspace{-3mm}
  \label{fig:delta_vs_oh2}
\end{figure}

In figure~\ref{fig:delta_vs_oh2} we show how the final DM relic density as well as the would-be scalar 
relic density (assuming no decay) scale with the mass splitting $\delta$. In the `forbidden' regime
with $m_\chi<m_\phi$ (and thus $\delta<0$), the relic density is as expected hardly affected by the presence of the 
dark-sector interactions $\chi\chi\leftrightarrow\phi\phi$: even for very small (positive) values of $\delta$, there
is simply not sufficient energy density stored in the scalars at SM freeze-out, cf.~eq.~(\ref{Yratioxpsi}), that
could subsequently be converted to DM particles. This abruptly changes as $m_\chi>m_\phi$, allowing DM
particles to annihilate very efficiently into $\phi$, after decoupling from the SM, 
while the reverse process is kinematically inhibited. This results in an exponential drop in the DM relic density for 
small negative values of $\delta$. The figure also shows that the dark sector decouples significantly later than
the SM in our scenario, $x_f^\phi\gg x_f^{\rm SM} \sim  12$, which justifies our treatment of the DM freeze-out
process as consisting of two distinct steps. As also expected, $\chi\chi\leftrightarrow\phi\phi$ remains in 
equilibrium most efficiently as $\delta\to 0$, in which case we can encounter freeze-out as late 
as $x_f^\phi\lesssim 10^4$ in the most extreme case.

We stress that the above derivation, and in particular the calculation of $\bar r$,
strongly relies on the fact that both $\chi$ and $\phi$ remain in kinetic equilibrium during the
freeze-out process. In fact, deviations of the velocity distributions from the Maxwellian
shape may in general strongly impact relic density calculations in `forbidden' DM
scenarios, making it necessary to move beyond Boltzmann equations based on number
densities alone~\cite{Binder:2021bmg, Aboubrahim:2023yag}. We estimate that kinetic
decoupling of $\chi$ from the SM happens approximately when electrons start to become
Boltzmann suppressed, i.e.\ $T \sim m_e$. This temperature is only slightly larger than the
decoupling temperature $T_f^\phi$, which is of the order of
$50\text{--}200 \, \mathrm{keV}$ (for the parameters of our best-fit regions). Concerning
the assumption that $\phi$ is kept in local equilibrium with $\chi$, we refer to the discussion
in the main text, cf.~footnote \ref{foot:phi_not_thermal}.

We also point out that a high-precision determination of the relic density would involve
coupled Boltzmann equations for the two abundances $Y_\chi$ and
$Y_\phi$~\cite{Bringmann:2020mgx}. While this would have the advantage of also being able to
treat dark Higgs boson decays, such a system of coupled Boltzmann equations would be
computationally much more expensive, and harder to smoothly integrate into the {\sf GAMBIT}
scanning framework. Our eq.~(\ref{eq:oh2final}), in contrast, presents an {\it estimate}
of the impact of $\phi\leftrightarrow \chi$ conversions after freeze-out from the SM that is both fast to
compute and sufficiently accurate for our purposes. 
The biggest concern in this respect is our rough estimate of the dark sector decoupling temperature,
cf.~eq.~(\ref{eq:Tdecphi}), combined with the fact that we find the final relic density to depend approximately 
linearly on the resulting value of $x_f^\phi$ for $0<\delta\ll1$. However, since the relic density depends 
exponentially on $\delta$ in this regime, as very clearly visible in figure \ref{fig:delta_vs_oh2}, 
even an $\mathcal{O}(1)$ uncertainty in the determination of $\Omega_\chi h^2$ would 
not have a drastic impact on the shape of the best-fit regions that we show in the main text.

\begin{figure}[t]
    \centering
    \includegraphics[width=0.49\linewidth]{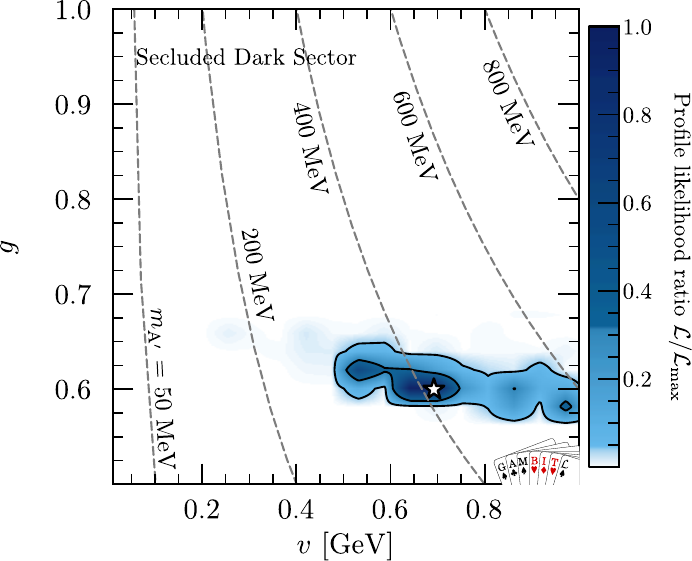}
    \hfill
    \includegraphics[width=0.49\linewidth]{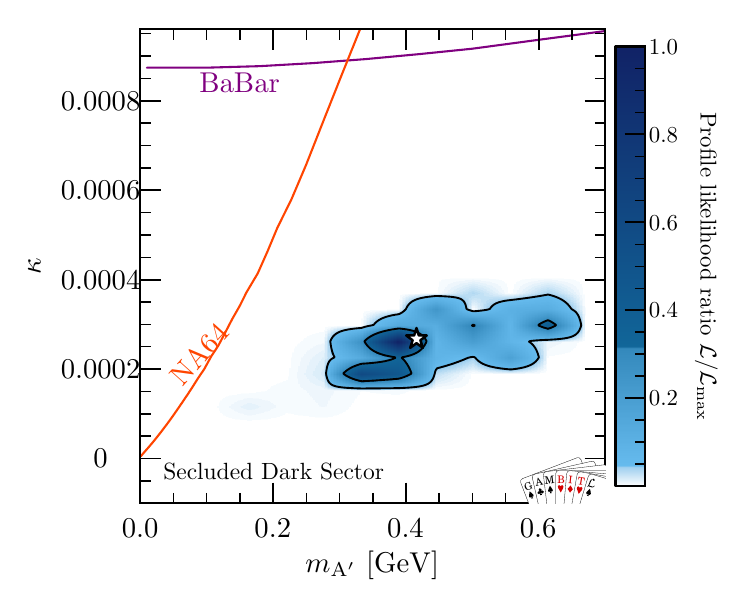}
    
    \includegraphics[width=0.49\linewidth]{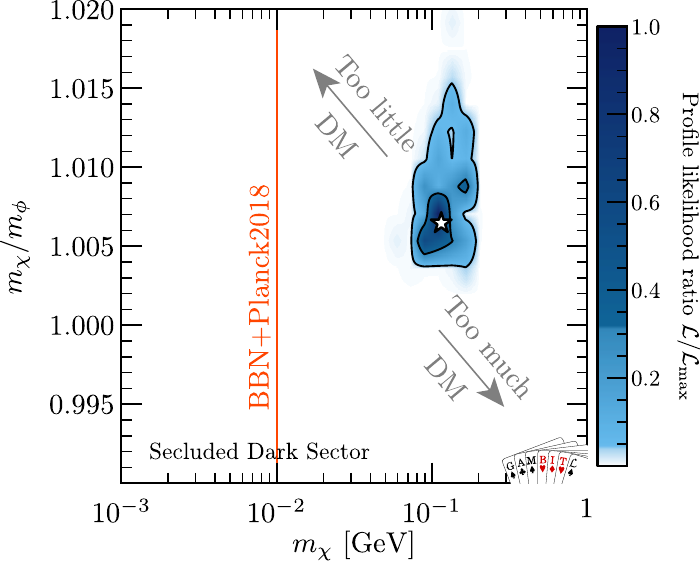}
    \hfill
    \includegraphics[width=0.49\linewidth]{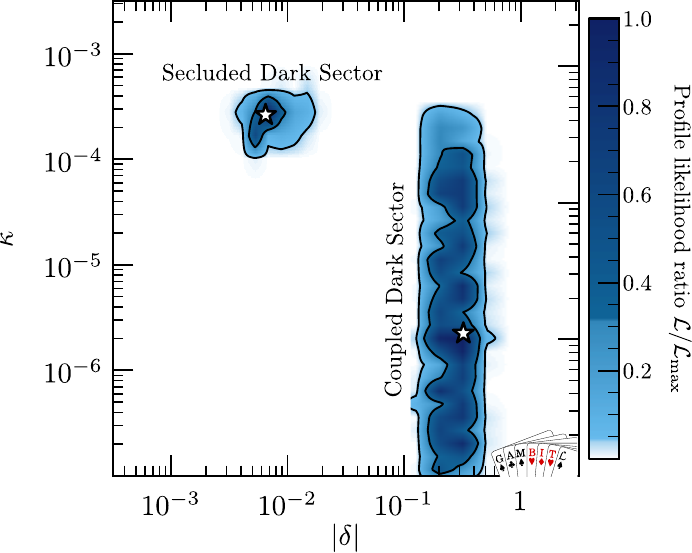}
    \caption{Same as figure~\ref{fig:secludedDS} in the main text, but including only points with $T_f^\phi > T_\text{decay}$, i.e.\ for which 
    the dark Higgs bosons decay only after DM decoupling.}
    \label{fig:secludedDS_trimmed}
\end{figure}

Related to this discussion, the above treatment of the dark sector relic density assumes that the dark Higgs
bosons do not decay before decoupling from the DM particles. The problematic parameter regions in this respect, 
with $T_f^\phi < T_\text{decay}$, are highlighted with a red color scale in the bottom-left panel of 
figure~\ref{fig:secludedDS-decays}. As already stressed in the main text, however, significant parts of our 
scan results do {\it not} face this problem. We show these parameter regions in figure~\ref{fig:secludedDS_trimmed}. 
This figure is the same as 
figure~\ref{fig:secludedDS}, except that all points with $T_f^\phi < T_\text{decay}$ have been removed. We find that this 
affects as expected
points with large $\kappa$, which correspond to larger $\Gamma_\phi$ and hence larger 
$T_\text{decay}$, as well as points with small $v$ and small $\delta$, which correspond to late decoupling and hence small 
$T_f^\phi$. As a result, the best-fit point shifts to larger vev and larger $\delta$, see point B in table~\ref{tab:bestfit}. 
Overall, we stress that we still find qualitatively very similar results compared to figure~\ref{fig:secludedDS}. The main 
difference is an apparently strongly constrained value of $\kappa\sim10^{-4}$ 
-- which however is just an artifact of the {\it ad hoc} procedure
of removing all parameter points with small dark Higgs boson lifetimes. In particular, the top-right panel of 
figure~\ref{fig:secludedDS_trimmed} does {\it not} imply that the secluded sector scenario cannot be probed by direct
searches for dark photons (see also section \ref{sec:outlook}).

We conclude this appendix by reporting, for completeness, the full annihilation cross sections
that enter in the relic density calculations discussed above:
\begin{align}
 (\sigma v)_{\chi\bar{\chi} \to e^{+}e^{-}} &= 
\frac{g^2 \kappa^2 e^2 (s-4 m_{\chi}^2) (2 m_{e}^2+s)}
{24 \pi s ((m_{A^{\prime}}^2-s)^2+\Gamma_{A^{\prime}}^{2}  m_{A^{\prime}}^2)} 
\frac{s}{2\left(s-2m_\chi^2\right)} \sqrt{1-\frac{4 m_{e}^2}{s}} \, . 
\end{align} 
\begin{align}  
   (\sigma v)_{\chi\bar{\chi} \to \phi\phi} &=
  \frac{y^{2}}{32 \pi  s v^2} \frac{s}{2\left(s-2m_\chi^2\right)} \sqrt{1-\frac{4 m_{\phi }^2}{s}} \\
  & \times \left[
    \frac{2 m_\phi^4 \left(s-4 m_{\chi }^2\right)}{\left(m_{\phi }^2-s\right){}^2}+\frac{8 m^2_{\chi } m_\phi^2}{m_{\phi }^2-s}
  +\frac{m_\chi^2 \left(m_{\phi }^2-4 m_{\chi }^2\right){}^2}{m_{\chi }^2 \left(4 m_{\phi }^2-s\right)-m_{\phi }^4}-2 m_\chi^2
    \right. \nonumber \\
  & ~~ +  \left( -\frac{2 m_\chi^2 \left(16 m_{\chi }^2 \left(m_{\phi }^2-s\right)+4 s m_{\phi }^2+32 m_{\chi }^4-6 m_{\phi }^4-s^2\right)}{\sqrt{s-4 m_{\chi }^2}
    \sqrt{s-4 m_{\phi }^2} \left(s-2 m_{\phi }^2\right)}
     \right. \nonumber\\
  & ~~ \left.\left.  +\frac{8 m_\chi^2 m_\phi^2 \left(8 m_{\chi }^2-2 m_{\phi }^2-s\right)}
    {\sqrt{s-4 m_{\chi }^2} \sqrt{s-4 m_{\phi }^2} \left(s-m_{\phi }^2\right)} \right)
    \tanh ^{-1}\left(\frac{\sqrt{s-4 m_{\chi }^2} \sqrt{s-4 m_{\phi }^2}}{s-2 m_{\phi }^2}\right)
    \right] \, . \nonumber\label{eq:cross-section-chichi-phiphi} 
\end{align}
Here, as in all expressions above, $v=\sqrt{s(s-4m_\chi^2)}/(s-2m_\chi^2)$ denotes as usual the velocity of one of the DM 
particles in the rest frame of the other.\footnote{For reference, the relative velocity of the DM particles
in the centre-of-mass frame is thus given by $2 v_{\rm CM}=2v(s-2m_\chi^2)/s$.}

\bibliographystyle{JHEP_improved}
\bibliography{literature}

\end{document}